\documentstyle[preprint,aps,epsf,prc]{revtex}
\begin{document}
\title{Electroproduction of pseudoscalar mesons on the deuteron.}
\author{Michail P. Rekalo \footnote{ Permanent address:
\it National Science Center KFTI, 310108 Kharkov, Ukraine}
}
\address{Middle East Technical University, 
Physics Department, Ankara 06531, Turkey}
\author{Egle Tomasi-Gustafsson}
\address{\it DAPNIA/SPhN, CEA/Saclay, 91191 Gif-sur-Yvette Cedex, 
France}
\author{Jacques Arvieux}
\address{\it Institut de Physique Nucl\'eaire, CNRS/IN2P3, 91406 Orsay Cedex 
France}
\maketitle
\date{\today}
\begin{abstract}
A general analysis of polarization phenomena for coherent meson 
electroproduction on deuterons, $e+\ d\to e+\ d\ +P^0$,  where $P^0$ 
is a pseudoscalar $\pi ^0$ or $\eta$-meson, is presented.
The spin structure of the electromagnetic current for $P^0$-production at 
threshold is parametrized in terms of specific (inelastic) threshold 
electromagnetic form factors which depend on the momentum transfer squared and 
the
effective mass of the produced hadronic system. We give expressions for the
structure functions of the reaction $e+\vec{d}\to e +d+ P^0$ 
(where the deuteron target is polarized) in terms of these
threshold form factors. The spin and isospin structures of the
$\gamma ^{*}+d\to d+P^0$ amplitudes (where $\gamma ^*$ is a virtual 
photon)
is established in the framework of 
the impulse approximation and relationships between meson electroproduction on 
deuterons and on nucleons are given. The reaction  of $\pi^0$ electroproduction 
on deuterons is investigated in detail both at threshold and in the region of 
$\Delta $-isobar excitation, using the effective Lagrangian approach for the 
calculation of the amplitudes of the elementary process $\gamma^*+N\to N+\pi$. 
Special attention is devoted to the analysis of all standard contributions to 
the exclusive cross section for $d(e,e\pi^0)d$, which are functions of the 
momentum 
transfer square, $k^2$, of the excitation energy of the produced hadrons and 
of the pion production angle, in a region of relatively large momentum transfer. 
The 
sensitivity of these contributions to different parametrizations of the  
$\gamma^*\pi\omega$ form 
factor as well as to the choice of $NN-$potential is discussed.

\end{abstract}
\pacs{20.25,25.30-c, 24.70+s, 13.40.Gp}

\section{Introduction}

The reaction $\gamma +d\to d+P^0$, where $P^0$ 
is a neutral pseudoscalar  meson ($\pi ^0$ or $\eta $), is the simplest 
coherent meson production process in $\gamma d$-collisions. The presence of 
a deuteron with zero isospin in the initial and final states leads to a specific 
isotopic structure for the corresponding amplitudes. 
Moreover, although the spin structure may be, in general, 
fairly complex, it is essentially simplified in the
near theshold region making the 
$\gamma +N\to N+P^0$ (where $N$ denotes a nucleon) and 
$\gamma +d\to d+P^0$ reactions especially interesting for hadron 
electrodynamics studies. 

The  $\gamma +d\to d+\pi ^0$ 
reaction is 
important to test the predictions of low energy theorems (LET) for threshold 
$\gamma +N\to N+\pi ^0$ amplitudes. Multipole analyses of 
older 
$\gamma +p\to p+\pi^0$ data \cite {Ma86,Be90} were in serious 
discrepancy with the predictions of LET \cite{Kr54,Am79}. Recent 
data\cite{Be96,Fu96} obtained with tagged photons, 
combined with new theoretical developments \cite{Ber96} have brought
experiment and theory into agreement. These calculations show that 
the 
amplitudes for the $\gamma +N\to N+\pi ^0$ reaction near threshold have 
a complex isotopic structure. Calculations of the electric dipole $E_{0+}$ 
threshold amplitudes for 
$\gamma +N\to N+\pi $ processes in the framework of the dispersion 
relation 
method \cite{Ha97} confirm this observation. Therefore the knowledge of the 
$\gamma +n\to n+\pi ^0$ reaction amplitude is very important and the 
$\gamma +d\to d+\pi ^0$ reaction appears the most 
suitable for that purpose. However the extraction of the  
$\gamma +n\to n+\pi ^0$ amplitude from $\gamma d$ experimental 
data \cite{Ar81} requires a careful study 
of possible rescattering effects \cite{Ko77,Bo78,Fa80}.

Pion-electropoduction $e+d\to e+d+\pi ^0$ is even richer since it 
involves 
longitudinal as well as transverse photons. Experimental information about this 
process has been missing for a long time, but such an experiment can be 
performed at MAMI \cite{Be96_2} or at Jefferson Lab. In this case,  with  the 
experimental set-up which has been used 
to measure the tensor deuteron polarization in elastic 
$ed$-scattering \cite{Ko94}, a sample of $\pi^0$-electroproduction data were 
obtained during dedicated runs, \cite{Abp99} at relatively large momentum 
transfer square  ($\simeq 1.1\div 1.6$ (GeV/c)$^2$) in the threshold and in the 
$\Delta$-region.

The $e+d\to e+d+\pi ^0$ reaction allows to "scan" the 
isospin 
structure in the full resonance region and to separate isovector from isoscalar 
contributions. Moreover, experiments using a polarized deuteron target yield a 
different information compared to measurements of the polarization of the final 
deuteron.

Another interesting problem of near threshold meson
photoproduction on deuterons concerns the isotopic structure of  the 
$\gamma+N\to S_{11}(1535)$ transition. The results of different multipole 
analyses of the 
$\gamma +N\to N+\pi $ reactions have shown that the 
$\gamma +N\to S_{11}(1535) $ transition is essentially 
isovector
\cite{Me74,Ar82,Cr83,Li93}, in agreement with predictions of quark models 
\cite{Fe71,Ko80,Li90,Wa90,Bi94}. Existing $\gamma +p\to p+\eta $
experimental data \cite{Kr95}   
in the near threshold region indicate that the 
$S_{11}( 1535) $ excitation is the main mechanism. 
On the other hand, the amplitude for $\gamma +d\to d+\eta $ in the
near threshold region has to be isoscalar and, therefore, small in contradiction 
with earlier 
data\cite{An69} which showed a large cross section. Recent 
$d( \gamma ,\eta ) $ $X$ data\cite{Kr95_2} have given an explanation
by showing that this reaction is essentially inelastic.

The near-threshold region for $\gamma \ +\ d\to d\ +\ \eta $ is
linked with the physics of the $n+p\to d+\eta $ reaction because both
processes are connected via the unitarity condition. The cross section 
for 
$n+p\to d+\eta $, which was first measured at Saturne \cite{Pl90}, was
found to be very large : $\sigma ( np\to d\eta ) =(100\pm 20) $ $\mu$b. 
A recent experiment at CELSIUS \cite{Ca97} has confirmed 
these data and has shown a steep decrease down to $\sigma\simeq 40$ $\mu$b up 
to  
$q_{CM}=20$ MeV, where $q_{CM}$ is the final kinetic energy in the reaction 
center 
of mass system (CMS). The shape of the energy dependence is reproduced by 
calculations 
taking into account the $N^*(1535)$ resonance \cite{La91,Mo96} but more exotic 
explanations are  not  ruled out\thinspace :

\begin{description}
\item[-]  the existence \cite{Ue91} of an isoscalar dibaryon resonance with
zero isotopic spin and a small width, $\Gamma\simeq 7$ $MeV$, or
\item[-]  the existence \cite{Wi96} of a quasi-bound $\eta d$-state (due to the
strong $\eta N$ interaction), or
\item[-] the possible presence of a 
nonperturbative $s\overline{s}$-component in the nucleon which could allow a 
strong 
$\eta $-production from spin singlet $np$ initial states \cite{Re97}.
\end{description}
The study of the processes $\gamma +d\to n+p+\eta$, $e+d\to 
e+d+\eta$ 
and $e+d\to  e+n+p+\eta$ with particular emphasis on polarization 
observables, 
would help to identify the correct interpretation.

We derive here a general analysis for pseudoscalar meson 
electroproduction on deuterons, based on general symmetry properties of the 
hadron 
electromagnetic interaction. A similar analysis limited to pion 
photoproduction on deuterons has been published \cite{St90,Re91,Re93}. Such a
general analysis has to be considered as the first necessary step in the 
theoretical study of this process and is no substitute for dynamical model 
calculations \cite{Ar00,Bea97}.
An adequate dynamical approach to pion electroproduction  has to take into 
account 
all previous theoretical findings related to other electromagnetic processes on 
deuteron, 
such as elastic $ed$ scattering \cite{Al99}, $\pi^0$-photoproduction,  
$\gamma+d\to d+\pi^0$ \cite{Me99}, 
and  deuteron photodisintegration $\gamma+d\to n+p$ \cite{Bo98}. Similarly to 
these 
processes, the reaction $e+d\to e+d+\pi^0$ will face two main problems: the 
study 
of 
the deuteron structure and of the reaction mechanism, on one side, and the 
determination of the neutron elementary amplitude 
($\pi^0$-meson electroproduction on neutron, $e^-+n\to e^-+n+\pi^0$), on another 
side.

Elastic $ed$-scattering, being the simplest process to access the deuteron 
structure, has been considered, for large momentum transfers,  a good case 
to test different predictions of perturbative QCD, such as the scaling  behavior 
of the deuteron electromagnetic form factors \cite{Ar75} and the 
hypothesis of helicity conservation \cite{Br73}. The analysis of the scaling 
behavior should help in defining the kinematical region of 
the transition regime from the meson-nucleon degrees of freedom to the 
quark-gluon description of the deuteron structure.  In this respect coherent 
$\pi^0$-electroproduction of the deuteron opens new possibilities to 
study the scaling phenomena in different regions due to the more 
flexible kinematical conditions: it unifies the kinematics of elastic 
$ed$-scattering, with its single dynamical variable (the momentum transfer 
square, 
$k^2$) and the process of $\pi^0$-photoproduction, with two independent 
dynamical variables (the total energy $s$ and the momentum transfer $t$ from the 
initial to the final deuteron). As a result, three kinematical variables  drive 
the process $e+d\to e+d+\pi^0$. Different mechanisms have a leading role 
in different kinematical regions. 
In order to interpret the first experimental data for $e+d\to 
e+d+\pi^0$, with small excitation energy of the produced $d\pi^0$-system (up to 
the $\Delta$-resonance region) but at relatively large momentum transfer, $k^2$, 
the starting point of the theoretical analysis is naturally the impulse 
approximation ({\it IA}). 
Similarly to  previous calculations of elastic $ed$-
scattering and $\pi^0$-photoproduction processes, as a 
further step, contributions of meson exchange currents (MEC) \cite{Ch74} have to 
be evaluated 
in the resonance region, while rescattering effects  \cite{Ko77,Bo78,Fa80} have 
to be taken into account in the near threshold region. Large disagreements 
exist, up to now, in a quantitative evaluation of these effects.

The present paper is organized as follows:

a) we first establish the spin structure of the matrix element for the 
$\gamma ^{*}+d\to d+P^0$ reaction and give a formalism for the description of 
polarization observables. 
The dependence of the $\vec{d}(e,e^{\prime }P^0)d$ differential 
cross section on the polarization characteristics of the
deuteron target is derived in a general form, using a formalism of
structure functions (SF), which is particularly adequate to describe, in the 
one-photon approximation, the polarization properties for any 
$e+A\to e+h+A^{\prime }$ process (where A is any nucleus and h
is a single hadron or hadronic system). These structure functions are further 
expressed in terms of the scalar amplitudes which parametrize the spin structure 
of the 
corresponding electromagnetic current for the process $\gamma^*+d\to d+P^0$.

b) we study the isospin structure of these reactions,

c) using the {\it IA} , we give relations between 
the scalar amplitudes, describing the $\gamma ^{*}+d\to d+P^0$  
and the $\gamma ^{*}+N\to N+P^0$ reactions,

d) we then examine the special kinematical conditions corresponding to threshold 
production,

e) finally, we calculate some observables for $e+d\rightarrow
e+d+\pi ^0$ in the framework of the {\it IA} in order to study its 
sensitivity to the isotopic structure of the $\gamma ^{*}+N\to N+\pi ^0$ 
processes near threshold and in the region of $\Delta $ excitation, at 
relatively large $-k^2$.

The present analysis has been extended to the electroproduction of a "scalar" 
deuteron ($i.e.$ $np$ pair with $J^P=0^+$) together with a pseudoscalar meson 
which 
would be much more difficult to investigate experimentally . The results are 
available on request to the authors.

\section{General formalism for the description of $\lowercase{e}+\ 
\lowercase{d}\to \lowercase{e}+\ \lowercase{d}\ +P^0$ processes}
\subsection{Derivation of the  cross section}
The general structure of the differential cross section for the 
 $e+\ d\to e+\ d\ +P^0$  reaction can be established in the 
framework of the one-photon mechanism (Fig. 1) by using only the most general 
symmetry properties of the hadronic electromagnetic interaction, such as gauge
invariance (the conservation of hadronic and leptonic electromagnetic
currents) and invariance upon mirror symmetry (parity invariance 
of the strong and electromagnetic interactions or, 
in short, $P$-invariance). 
The details of the reaction mechanism
and the deuteron structure do not contribute at this step.

The transition matrix element can be written:
\begin{eqnarray}
\mathcal{M}( ed\to edP) &=&\frac{e^2}{k^2}\overline{u}(k_2)
\gamma _\mu u(k_1) \left\langle dP\left| 
\hat{\mathcal{J}_\mu }\right| d\right\rangle \equiv \frac{e^2}{k^2}\ell
_\mu \mathcal{J}_{\mu,}  \label{li5} \\
\ell _\mu &\equiv &\overline{u}(k_2) \gamma _\mu u(k_1) ,\mathcal{J}_\mu \equiv 
\left\langle dP\left| \hat{\mathcal{J}_\mu }\right| d\right\rangle,  \nonumber
\end{eqnarray}
where the notations of the particle four-momenta are explained in Fig. 1 and 
$\mathcal{J}_\mu $ is the electromagnetic current for the transition 
$\gamma^{*}+d\to d+P^0$.
Using the conservation of leptonic and hadronic currents, 
($k\cdot \mathcal{J}=k\cdot \ell =0$) one can rewrite the matrix element in 
terms of space-like components of currents only :
\[
\mathcal{M}=\frac{e^2}{k^2}\vec{e}\cdot \vec{\mathcal{J}},
\vec{e}\equiv \vec{\ell }-\vec k\frac{      
\vec k\cdot \vec{\ell }}{{k_0}^2}, 
\]
where $k=({k_0}, \vec k)$, ${k_0}$ is the energy,
${\vec k}$ is the three-momentum of the virtual photon in the 
CMS of $\gamma^*+d\to d+P^0$. All observables  
will be determined by bilinear combinations of the components of the hadronic 
current
$\vec{\mathcal{J}}$: $H_{ab}=\mathcal{J}_a\mathcal{J}_b^{*}$. As a result, we 
obtain the following formula for the exclusive differential cross section
in terms of the tensor components $H_{ab}$:
$$
\frac{d^3\sigma }{dE_2 d\Omega_e d\Omega_p} =\frac{\alpha ^2}
{64\pi ^3}\frac{E_2}{E_1}\frac{\left| \vec q\right| }{M\sqrt{s}}\frac
1{1-\kappa }\frac{\emph{X}}{( -k^2) },  
$$
$$
\emph{X} =H_{xx}+H_{yy}+\kappa \cos 2\varphi \left( H_{xx}-H_{yy}\right)
-2\kappa \frac{k^2}{{k_0}^2}H_{zz}
$$
\begin{equation}
-\sqrt{2\kappa ( 1+\kappa ) \frac{( -k^2) }
{{k_0}^2}}
\left[ \cos \varphi \left( H_{xz}+H_{zx}\right) -\sin \varphi \left(
H_{yz}+H_{zy}\right) \right] 
\end{equation}
$$
+\kappa \sin 2\varphi \left( H_{xy}+H_{yx}\right) -\lambda \sqrt{1-\kappa 
}\Large[ \sqrt{1+\kappa }\left( H_{xy}-H_{yx}\right) -
\sqrt{2\kappa \frac{( -k^2) }{{k_0}^2}} 
$$
$$ 
( \sin \varphi ( H_{xz}-H_{zx}) -\cos \varphi
( H_{yz}-H_{zy}) 
\Large ], 
$$
where $\kappa ^{-1}=1-2\vec k_L^2tg^2\displaystyle\frac{\theta e}2/k^2$ is the 
polarization of the virtual photon.
Here $E_1( E_2) $ is the energy of the initial (final) electron
in the lab system; $\theta _e$ is the electron scattering angle in the lab; 
$d\Omega _e$ is the solid angle of the scattered electron in the lab system ; 
$d\Omega _p$ and $\vec q$ are respectively the solid angle and three-momentum 
of 
the  produced $P^0$-meson in the $CMS$; $M$ is the target mass; 
$\vec k_L$ is the photon three-momentum 
in the lab system; $\lambda =\pm \ 1$ for the two possible initial electron 
helicities; $\varphi $ is the azimuthal angle of the
scattered electron with respect to the plane of the reaction $\gamma^*+d\to 
d+P^0$. The coordinate system is such that
the $z$-axis is along $\vec k$ and the $xz$ plane is defined by
$\vec k$ and $\vec q$.

The tensor structure of $H_{ab}=\overline{\mathcal{J}_a\mathcal{J}_b^{*}}$ 
(where the line denotes the sum over the final deuteron polarizations) 
can be written in the following form:
\begin{equation}
H_{ab}=H_{ab}^{( 0) }+H_{ab}^{( 1) }+H_{ab}^{(
2) }, \label{li17}
\end{equation}
\noindent where the indexes $(0) $, $(1) $ and $(2) $ correspond to unpolarized, 
vector and
tensor polarized initial deuterons, respectively. The first term $H_{ab}^{(0)}$ 
can be parametrized as:
\begin{equation}
H_{ab}^{(0) } = \hat m_a\hat m_bh_1+\hat n_a\hat 
n_bh_2+\hat k_a\hat k_bh_3+
\left\{ \hat m,\hat k\right\} _{ab}h_4+i\left[ \hat m,\hat 
k\right] _{ab}h_5, 
\end{equation}
with $\left\{ \hat m,\hat k\right\} _{ab} =\hat m_a\hat k_b+\hat 
m_b\hat k_a, ~~
\left[ \hat m,\hat k\right] _{ab} =\hat m_a\hat k_b-\hat 
m_b\hat k_a. $
Here $h_1$ - $h_5$ are the real SF's, which depend on $k^2$, $s$ and $t$, 
$\hat {\vec{n}} = \vec k\times  \vec q/\left| 
\vec k\times \vec q\right|$, $\hat {\vec {m}} =\hat {\vec {n}}\times 
\hat{\vec k}$, $\hat{\vec k}=\vec k/|\vec k|$.
The SF's $h_1$ - $h_4$ determine the cross section for the reaction
$e+d\to e+d+P^0$  with unpolarized 
particles. The SF $h_5$ (the so-called ``fifth'' structure function) 
determines the asymmetry of longitudinally polarized electrons scattered 
by an unpolarized target. This $T$-odd contribution is determined by the 
product of longitudinal and transverse components of the hadron electromagnetic 
current and it is
nonzero only for noncoplanar kinematics, $\varphi \neq 0$. This
contribution is very sensitive to the details of the final state interaction.

The tensor $H_{ab}^{(1)}$ is linear in the pseudovector 
$\vec{S}$ (vector polarization of the initial deuteron) and can be written
in the following general form:
$$
H_{ab}^{(1)} =\hat {\vec m}\cdot \vec{S}(
\{ \hat m,\hat n\} _{ab}h_6+\{ \hat k,\hat n\}_{ab}h_7+i[ 
\hat m,\hat n]
_{ab}h_8+i[ \hat k,\hat n] _{ab}h_9) 
$$
$$
 +\hat {\vec n}\cdot       
\vec{S}( \hat m_a\hat m_bh_{10}+\hat n_a\hat n_bh_{11}+
\hat k_a\hat k_bh_{12}+\{\hat  m,\hat k\} _{ab}h_{13}+
i[ \hat  m,\hat  k]_{ab}h_{14}) 
$$
\begin{equation}
+\hat{\vec k}\cdot \vec{S}( \{ \hat m,\hat n\}
_{ab}h_{15}+\{ \hat k,\hat n\} _{ab}h_{16}+i[ \hat m,\hat n]
_{ab}h_{17}+i[ \hat k,\hat n] _{ab}h_{18}). 
\end{equation}
So, 13 real SF's $h_6$ - $h_{18}$ describe the effects of the vector target 
polarization for the exclusive cross section in the one-photon approximation. 
The symmetric (antisymmetric) part of $H_{ab}^{(1)}$ 
determines the scattering of unpolarized (polarized)
electrons by a vector-polarized target. In particular, it is the symmetric 
part of $H_{ab}^{( 1) }$, which induces $T$-odd asymmetries in the 
$\vec{d}(e,e^{^{\prime }}P^0)d$ reaction.

The integration of the tensor $H_{ab}^{(1)}$ over $d\Omega{_p}$ can be done in 
the following way, typical for inclusive
polarized electron-hadron collisions \footnote {Note, that for an unpolarized 
deuteron target the following formula holds:
$\int H_{ab}^{(0)}d\Omega_p=
\delta_{ab}w_1+\hat {k_a}\hat{k_b}w_2$.}:
$$
\int H_{ab}^{(1) }d\Omega {_p} =i\varepsilon _{abc}S_c w      
_3+i\varepsilon _{abc}\hat k_c\vec{S}\cdot \hat {\vec k}w_4+\left( 
\hat{ k_a}\left
[ \hat {\vec k}\times \vec{S}\right]_b+\hat {k_b}\left[ 
\hat {\vec k}\times \vec{S}\right] 
_a\right) w_5.
$$
For the inclusive structure functions $w_3$ - $w_5$ one obtains the 
following expressions in terms of integrals of the linear combinations of SF's 
$h_i$:
\begin{eqnarray}
w_3 &=&\int (-h_{9}-h_{14}+h_{17})d\Omega _p,  \nonumber \\
w_3+w_4 &=&\int  h_{17}d\Omega _p,  \label{li20} \\
w_5 &=&\int (h_7-h_{13})d\Omega _p , \nonumber
\end{eqnarray}
$i.\ e.$ most of the exclusive $SF^{^{\prime }}s$ $h_6$ - $h_{18}$
do not contribute to  the inclusive $SF^{^{\prime }}s$ $
w_3$ - $w_5$.

Finally, for the tensor $H_{ab}^{(2) }$, characterizing the
effects of the tensor target polarization, it is possible to write the
following general expression :
\begin{eqnarray}
H_{ab}^{( 2) } &=&( Q_{cd}\hat m_c\hat m_d) (
\hat m_a\hat m_bh_{19}+\hat n_a\hat n_bh_{20}+\hat 
k_a\hat k_bh_{21} +\{ \hat m,\hat k\} _{ab}h_{22}+i[ \hat 
m,\hat k]
_{ab}h_{23})   \nonumber \\
&&+( Q_{cd}\hat n_c\hat n_d) (
\hat m_a\hat m_bh_{24}+\hat n_a\hat n_bh_{25}+\hat 
k_a\hat k_bh_{26} +\{ \hat m,\hat k\} _{ab}h_{27}+i[ \hat 
m,\hat k]
_{ab}h_{28})   \nonumber \\
&&+( Q_{cd}\hat m_c\hat k_d) (
\hat m_a\hat m_bh_{29}+\hat n_a\hat n_bh_{30}+\hat 
k_a\hat k_bh_{31} +\{ \hat m,\hat k\} _{ab}h_{32}+i[\hat  
m,\hat k]
_{ab}h_{33})   \\
&&+( Q_{cd}\hat m_c\hat n_d) ( \{ \hat m,\hat n\} _{ab}h_{34}+
\{
\hat k,\hat n\} _{ab}h_{35} +i[ \hat m,\hat n] _{ab}h_{36}+i[ 
\hat k,\hat n]
_{ab}h_{37})   \nonumber \\
&&+( Q_{cd}\hat k_c\hat n_d) ( \{ \hat m,\hat n\}_{ab}h_{38}
+ \{\hat k,\hat n \}_{ab}h_{39} 
+i[ \hat  m, \hat n ]_{ab}h_{40}+i[ \hat k,\hat n ]_{ab} h_{41}),  \nonumber
\end{eqnarray}
where $Q_{ij}$ is a tensor polarization component of the deuteron target,
$Q_{ii}=0$, $Q_{ij}=Q_{ji}$, so the density matrix for the initial 
deuteron can be written as follows:
\begin{equation}
D_{1a}D_{1b}^{*}=\frac 13\left( \delta _{ab}- \frac 32 i\varepsilon
_{abc}S_c-Q_{ab}\right).
\end{equation}
 
Therefore, for exclusive reactions like $A(e,e)A^{\prime }h$, in 
the framework of the one-photon mechanism, the 
effects of the target 
tensor polarization are characterized  by a set of 23 real SF's, 
$h_{19}$ - $h_{41}$. However the result of the integration of this tensor 
over the angle $d\Omega _p$ of the $P^0$-meson reduces its dependence to 5 
real structure functions only :
\begin{eqnarray*}
\int H_{ab}^{(2}d\Omega _p &=&\left( Q_{cd}\hat k_c\hat 
k_d\right)
\left[ w_6\left( \delta _{ab}-\hat k_a\hat k_b\right) +w      
_7\hat k_a\hat k_b\right]  \\
&&+Q_{ab}w_8+\left( Q_a\hat k_b+Q_b\hat k_a\right) w_9+i\left(
Q_a\hat k_b-Q_b\hat k_a\right) w_{10},~Q_a=Q_{ab}\hat k_b.
\end{eqnarray*}

In summary, the exclusive differential cross section for unpolarized electron
scattering  in $ e^-+d\to e^-+d+P$ is determined by a set of 28
($4_0+8_1+16_2=28$) SF's, where the indexes $0$, $1$ and 
$2$ correspond to unpolarized target $(0) $, target with vector 
$(1) $ and tensor $(2)$ polarizations. For 
longitudinally polarized electron scattering there are additional $
1_0+5_1+7_2=13$ SF's.
These 41 SF's can be divided alternatively into 5 - describing electron 
scattering 
by an unpolarized deuteron target, 13 - describing the effect of the vector 
deuteron polarization and 23 - depending on the tensor deuteron polarization. 
Taking into account the T-invariance of the electromagnetic interaction of 
hadrons, we can classify the set of 41 SF's in $1_0+8_1+7_2=16$ T-odd 
structures and $4_0+5_1+16_2=25$ T-even SF, as illustrated in Table 1.

For inclusive hadron electro-production, the number of SF's reduces to two 
$(w_1-w_2)$ for the unpolarized case, three $(w_3-w_5)$, describing deuteron 
vector polarization effects and five $(w_6-w_{10})$, depending on the tensor 
polarization. 

This analysis takes into account the eventual vector and tensor
polarizations of the target but not the polarization of the produced particles
since a summation over the final polarization states has been done. It can be 
easily generalized  to any other polarization observables such as the recoil 
deuteron polarization  or the spin correlation coefficients.

\subsection{Amplitude analysis}

The next step in this analysis, is to establish the spin structure of the matrix 
element for the $\gamma ^{*}+d\to d+P^0$ reaction without 
any constraint on the kinematical conditions.

This spin structure of the amplitude can be parametrized by different 
(and equivalent) methods, but  for the analysis of polarization 
phenomena the choice of {\it transverse amplitudes} is sometimes  preferable.
Taking into account the $P$-invariance of the electromagnetic interaction of
hadrons, the dependence of the amplitude of  $\gamma ^{*}+d\to d+P^0$ on 
the $\gamma ^{*}$ polarization and polarization three-vectors $\vec{D_1}$ and 
$\vec{D_2}$ of the initial and final 
deuterons is given by:
$$
F( \gamma ^{*}d\to dP^0) =\vec{e}\cdot\hat {\vec m}
( g_1\hat {\vec {m}}\cdot \vec{D_1}
\hat {\vec n}\cdot \vec{D_2^{*}}+g_2\hat {\vec k}\cdot 
\vec{D_1}\hat {\vec n}\cdot \vec{D_2^{*}}
+g_3\hat {\vec n}\cdot \vec{D_1}\hat {\vec m}\cdot 
\vec{D_2^{*}}+g_4\hat {\vec n}\cdot \vec{D_1}
\hat {\vec k}\cdot \vec{D_2}^{*})$$
$$
+\vec{e}\cdot \hat{\vec n}( g_5\hat{\vec m}\cdot 
\vec{D_1}\hat {\vec m}\cdot \vec{D_2^{*}}+g_6      
\hat {\vec n}\cdot \vec{D_1}\hat {\vec n}\cdot \vec{
D_2^{*}}  +g_7\hat {\vec k}\cdot \vec{D_1}\hat {\vec k}\cdot 
\vec{D_2^{*}}+g_8\hat {\vec {m}}\cdot \vec{D_1}
\hat {\vec k}\cdot \vec{D_2^{*}}+g_9\hat {\vec k}\cdot 
\vec{D_1}\hat {\vec {m}}\cdot \vec{D_2^{*}})  
$$
\begin{equation}
+\vec{e}\cdot \hat {\vec k}( g_{10}\hat {\vec {m}}\cdot 
\vec{D_1}\hat {\vec n}\cdot \vec{D_2^{*}}+g_{11}
\hat {\vec k}\cdot \vec{D_1}\hat {\vec n}\cdot \vec{      
D_2^{*}}   
+g_{12}\hat {\vec n}\cdot \vec{D_1}\hat {\vec {m}}\cdot 
\vec{D_2^{*}}+g_{13}\hat {\vec n}\cdot \vec{D_1}
\hat {\vec k}\cdot \vec{D_2^{*}}), \label{li22} 
\end{equation}

The process  $\gamma ^{*}+d\to d+P^0$ is described in the general case, 
by a set of 9 amplitudes for the absorption of a virtual photon with transverse 
polarization and 4 amplitudes for the absorption of a virtual photon with 
longitudinal polarization. These numbers are dictated by the values of the spins 
of the particles and by the P-invariance of hadron electrodynamics. Taking into 
account the possible helicities for $\gamma^*$ and deuterons ( in the initial 
and final states) one can find $3~(\gamma^*)\times 3~(initial~ deuteron)
\times 3~(final~ deuteron)=27$ different transitions in $\gamma ^{*}+d\to d+P^0$ 
and 27 corresponding helicity amplitudes 
$f_{\lambda_\gamma;\lambda_1;\lambda_2}$, where 
$\lambda_i$ are the corresponding helicities. Not all these amplitudes are 
independent, due to the following relations:
$f_{-\lambda_\gamma;-\lambda_1;-\lambda_2}=-
(-1)^{\lambda_\gamma-\lambda_1-\lambda_2}
f_{\lambda_\gamma;\lambda_1;\lambda_2}$, which result from the P-invariance. It 
is then possible to find that 
$f_{00,0}=0$ and that it  remains only  13 independent complex  amplitudes.
Therefore the  complete experiment requires, at least, the 
measurement of 25 observables. Let us mention in this respect specific 
properties of polarization phenomena for inelastic electron-hadron scattering: 
in exclusive $e+d\to e+d+P^0$ processes the virtual photon  
has a nonzero linear polarization, even for the scattering of unpolarized 
electrons by an unpolarized deuteron target. Therefore, the study of the 
$\varphi$- and $\kappa$-dependences of the $d(e,e P^0)d$
differential cross section - at a fixed values of the dynamical variables 
$s$, $t$ and $k^2$ - allows, in principle, to find not a single, but 4 different 
quadratic combinations of scalar amplitudes simultaneously:
$h_1$, $h_2$, $h_3$ and $h_4$.
The relationships between the structure functions $h_i,i=1-41$, and the 
amplitudes 
$g_k,k=1-13$, are given in the Appendix.

\section {The $\gamma ^{*}+\lowercase{d}\to \lowercase{d}+P^0$ reaction at 
threshold}

\subsection{Derivation of the cross section}

The threshold region is defined here as the $\gamma^*$ energy region in which  
$P^0$-meson production occurs in a S-state. This region may be wide as it
happens in $\gamma +N\to N+\eta $ or very narrow as in 
$\gamma +p\to p+\pi ^0$. This region starts from
$s=\left( M+m_{P}\right) ^2$, where $m_{P}$ is the mass of the produced 
pseudoscalar meson,  but the momentum transfer squared $k^2$ can take 
any value in the space-like region $(k^2\leq 0)$.

For threshold $P^0$-meson production only one
three-momentum, $\vec k$, is present (instead of two: $\vec k$ and $\vec q$, 
in the general case) and the full
kinematics of the produced hadronic system is fixed by the kinematical
conditions of the scattered electron only, similarly to  elastic
$ed$-scattering. For S-wave production any angular dependence 
in $\gamma^{\ast} +d\to d+\pi ^0$
disappears and the corresponding integration can be done
trivially : $\int X^{(t)}d\Omega _p=4\pi X^{(t)}. $
Setting $\varphi =0$ means that for inclusive electron
scattering, the $xz$ plane is related to the electron scattering plane. 
The inclusive cross section is obtained by integrating the differential 
cross section (2):
$$
\frac{d^2\sigma }{dE_2d\Omega _e} =\frac{\alpha ^2}{16\pi ^2}\frac{E_2}{E_1      
}\frac{\left| \vec q\right| }{M\sqrt{s}}\frac 1{1-\kappa }
\frac{X^{(t)}}{(-k^2) }, $$
$$X^{(t)} =H^{(t)}_{xx}+H^{(t)}_{yy}+\kappa 
\left(H^{(t)}_{xx}-H^{(t)}_{yy}\right)
-2\kappa \frac{k^2}{{k_0}^2}H^{(t)}_{zz}-\sqrt{2\kappa 
\left( 1+\kappa \right) \frac{(-k^2)}{{k_0}^2}}
\left( H^{(t)}_{xz}+H^{(t)}_{zx}\right) $$
\begin{equation}
-\lambda \sqrt{1-\kappa }\left( \sqrt{1+\kappa }
\left( H^{(t)}_{xy}-H^{(t)}_{yx}\right) +      
\sqrt{2\kappa \frac{(-k^2) }{{k_0}^2}}
\left( H^{(t)}_{yz}-H^{(t)}_{zy}\right)
\right), \label{li7} 
\end{equation}
where the superscript $(t)$ stands for threshold.

The hadronic tensor $H_{ab}^{(t)}$, for the case of polarized deuteron target, 
can be written as :
$$
H_{ab}^{(t)} =\left( \delta _{ab}-\hat{ k_a}\hat{ k_b}\right) 
t_1\left(k^2\right) +\hat{ k_a}\hat{ k_b}\ t_2(k^2) 
+i\varepsilon _{abc} S_c t_3(k^2) +i\varepsilon
_{abc}\hat{ k_c}\vec{S}\cdot \vec kt_4(k^2) 
$$
$$
+\left[ \hat{ k_a}\left( \vec k\times \vec{S}\right)
_b+\hat{ k_b}\left( \vec k\times \vec{S}\right) _a\right]
t_5(k^2) 
+\left( \vec Q \cdot \hat {\vec k}\right) \left[ \left( \delta _{ab}-\hat{ 
k_a}\hat{ k_b}\right)
t_6(k^2) +\hat{ k_a}\hat{ k_b}t_7(k^2) \right]  $$
\begin{equation}+Q_{ab}t_8(k^2) +\left( Q_a\hat{ k_b}+Q_b\hat{ k_a}\right) 
t_9\left(
k^2\right) 
+i\left( Q_a\hat{ k_b}-Q_b\hat{ k_a}\right) t_{10}(k^2). \label{li8} 
\end{equation}

The quantities $t_i(k^2) $, $i=1$ - $10$, are real structure 
functions, which are bilinear combinations of threshold electromagnetic
form factors which will be defined in the next section.

The symmetrical part of the tensor $H_{ab}^{(t)}$ determines the
differential threshold cross section for the scattering of unpolarized electrons
(by polarized and unpolarized deuterons), and the antisymmetrical part
characterizes the scattering of longitudinally polarized electrons.

\subsection{Amplitude analysis }

Taking into account the $P$-invariance of the hadronic electromagnetic
interaction,  the following threshold multipole transitions for 
$\gamma^{*}+d\to d+P^0$ are allowed:
$$
E1_\ell ,E1_t\text{ and }M2\to \mathcal{J}^P=1^{-}, 
$$
where $\mathcal{J}$ and $P$ are respectively the total angular momentum and 
parity of the $\gamma ^*d$ system.
Therefore, threshold $P^0$-electroproduction is characterized by 
two transitions with absorption of electric dipole virtual photons
(with longitudinal $ \ell $ and transverse 
$t $ polarizations) and one transition with absorption of 
magnetic quadrupole (transverse only) virtual photons.

The threshold amplitude of the process $\gamma ^{*}+d\to d+P^0$ can be 
parametrized in the following way:
\begin{eqnarray}
F_{th} &=&\left( \vec{e}\cdot \vec{D_1}\times 
\vec{D_2^{*}}-\vec{e}\cdot \hat {\vec k}\ 
\vec{D_1}\times \vec{D_2^{*}}\cdot \hat {\vec k}
\right) f_{1t}(k^2)  \label{li4} \\
&&+\vec{e}\cdot \hat {\vec k}\ \vec{D_1}\times 
\vec{D_2^{*}}\cdot \hat {\vec k}f_{1l}(k^2)   
\nonumber \\
&&+\left( \vec{e}\times \hat {\vec k}\cdot \vec{D_1}\ 
\hat {\vec k}\vec{\cdot D_2^{*}}+\vec{e}\times 
\hat {\vec k}\vec{\cdot D_2^{*}}\ \hat {\vec k}\cdot       
\vec{D_1}\right) f_2(k^2),  \nonumber
\end{eqnarray}
where $\vec{e}$ is the polarization of the virtual $
\gamma $-quantum.

The form factor $f_{1t}(k^2) \left[ f_{1\ell }(k^2)
\right] $ describes the absorption of electric dipole virtual photons with 
transverse $[$longitudinal$]$ 
polarization and the form factor $f_2(k^2) $, the absorption 
of a magnetic quadrupole $\gamma $-quantum. They have the same fundamental
meaning as the elastic electromagnetic form factors of the deuteron. 

Generally they are complex functions of $k^2$, due to the unitarity 
condition (Fig. 2) in the variable $s$  ( with a $n+p$ system in an 
intermediate state with both nucleons on the mass shell). 
But their relative phases have to be  equal to $0$ or $\pi $,  
as a result of $T$-invariance of
hadron electrodynamics (theorem of Christ and Lee \cite{Ch66}).
In general, they depend also on the $s$ variable, 
so that $f_i(k^2) \to f_i( k^2,s) $. 

In order to have a full reconstruction of the spin structure  for 
$\gamma ^{*}\,+\,d\to d\,+\,P^0$, polarization 
measurements are necessary. A simple one is  the tensor polarization of 
the scattered deuteron (or the tensor analyzing power using a  polarized
deuteron target).

After summing over the polarization states of the final deuterons the following
expressions can be obtained for the threshold $SF^{^{\prime }}s\ t_1$ - 
$t_{10}$ in terms of the electromagnetic threshold form factors 
$f_{1t}(k^2) $, $f_{1\ell }(k^2) $ and 
$f_2(k^2) $\quad :
$$
3t_1(k^2) =2\left( \left| f_{1t}(k^2) \right|
^2+\left| f_2(k^2) \right| ^2\right),  $$
$$
3t_2(k^2) =2\left| f_{1\ell }(k^2) \right| ^2\text{      
,}  $$
$$
t_3(k^2) =-\frac 12\cal{R}e~f_{1\ell }(k^2)\left(
f_{1t}(k^2)+f_2(k^2)\right) ^{*},  $$
$$
t_4(k^2) =-\frac 12\left| f_{1t}(k^2)-f_2\left
( k^2\right)\right| ^2+\frac 12      
\cal{R}e~f_{1\ell }\left( f_{1t}(k^2)+f_2(k^2)\right) 
^{*}, $$
$$
t_5(k^2) =-\frac 12\cal{I}m~f_{1\ell }(k^2)\left(
f_{1t}(k^2)+f_2(k^2)\right) ^{*},  $$
$$
3t_6(k^2) =-4\cal{R}e~f_{1t}(k^2)f_2^{*}\left
( k^2\right), $$
$$
3t_7(k^2) =\left| f_{1t}(k^2)-f_2(k^2)
\right| ^2-2\cal{R}e~f_{1\ell
}(k^2)\left( f_{1t}(k^2)+f_2(k^2)\right) 
^{*},  $$
$$
3t_8(k^2) =\left| f_{1t}(k^2)-f_2(k^2)
\right| ^2, $$
$$
3t_9(k^2) =-\left| f_{1t}(k^2)-f_2(k^2)
\right| ^2+\cal{R}e~f_{1\ell
}(k^2)\left( f_{1t}(k^2)+f_2(k^2)\right) 
^{*}
$$
\begin{equation}
3t_{10}(k^2) =-\cal{I}m~f_{1\ell }(k^2)
\left( f_{1t}(k^2)+f_2(k^2)\right)
^{*}\text{.}  \label{li11}
\end{equation}
This a strong simplification compared to the 41 real SF's, depending on 13 
complex amplitudes, which are necessary in the general case.

The $SF$ $t_5(k^2)$ is related to the asymmetry of unpolarized 
electrons scattered by a vector polarized deuteron target (with polarization
orthogonal to the electron scattering plane), while the SF 
$h_{10}(k^2)$ is related to the asymmetry of longitudinally 
polarized electrons scattered by a deuteron target with tensor polarization. 
These two $SF^{^{\prime }}s$ are determined by the interference of the 
longitudinal $\left(f_{1\ell }(k^2) \right) $ and both transverse 
($f_{1t}$\thinspace and $f_2$) form factors of the threshold transition 
$\gamma^{*}+d\to d+P^0$. They define the T-odd polarization 
observables and must vanish if the relative phase
of the longitudinal and transverse form factors is equal to $0$ or $\pi $
\cite{Ch66}. A dedicated experiment at SLAC \cite{Ro70} for the
search of T-odd asymmetry of unpolarized electrons (and positrons) by a
polarized proton target - with negative result - remains the
best test of T-invariance in hadron electrodynamics (at moderate energies). 
No similar experiments have been done with a polarized deuteron target but 
an attempt \cite{Pr68} to detect a nonzero vector deuteron polarization in 
elastic $ed$-scattering has been tried, with a negative result too.

From the expressions, obtained for the $SF^{^{\prime }}s$ in terms of the
corresponding threshold form factors, one can find an optimal strategy for 
performing a full experiment on  $P^0$-meson electroproduction on deuteron 
near threshold. One must first perform a  Rosenbluth separation
for the differential cross section of unpolarized electron scattering by an
unpolarized target, which allows to find the structure functions 
$t_1(k^2) $ and $t_2(k^2)$. These $SF^{^{\prime }}s$
determine the total cross sections for the absorption of virtual photons with 
transverse and longitudinal polarizations. 
It is straightforward then to deduce, from the
longitudinal structure function $t_2(k^2)$,  the 
$k^2$-dependence of the form factor $f_{1\ell }(k^2)$ 
- for absorption of electric dipole longitudinal virtual
photons. 

The transverse structure function $t_1(k^2) $ contains
the contributions of both transverse electromagnetic form factors, namely $
|f_{1t}| ^2$ and $| f_2(k^2) | ^2$. 
If we interchange the transverse and longitudinal structure functions, we 
have a situation similar to elastic $ed$-scattering : for 
elastic $ed$-scattering the transverse structure function contains
only the contribution of the magnetic form factor, so its $k^2$-dependence 
can be found directly (after a Rosenbluth fit), but the longitudinal 
structure function contains the contributions of the charge and quadrupole
electromagnetic form factors of deuteron. To separate these contributions it 
is necessary to measure the tensor polarization of scattered deuterons or the 
tensor analyzing power \cite{Ca16}. From this we can conclude that the 
measurement of the tensor polarization of the final deuteron in 
$e+d\to e+\,\,d\,+\,P^0$\ near threshold, will allow to 
separate the contributions due to $f_{1t}( k^2) $ and $ f_2(k^2)$.

This procedure, however,  does not give the sign of the threshold form factors. 
For elastic $ed$-scattering, using the well 
known values of the static electromagnetic characteristics of the deuteron : 
its electric charge, magnetic and quadrupole moments, it is possible to
extrapolate the sign step by step for any values of the momentum transfer 
square $k^2$. We can use the same method for $\gamma^{*}+d\to d+P^0$, 
using at $k^2=0$ the signs of the amplitude for 
$\gamma\,+\,d\to d\,+\,P^0$ which can be deduced, in principle, 
from the signs of the threshold amplitudes for the elementary processes 
$\gamma+N\to N+P^0$. 

We can also find the sign of the $f_{1t}(k^2)$, $f_{1\ell }(k^2)$ and 
$f_2(k^2)$ form factors at any value $k^2$ by using their relation with the
form factors of the  $\gamma^{*}+N\to N+P^0$ reactions at threshold.
The matrix element for S-wave
$P^0$-meson production on a nucleon can be parametrized in terms of two form
factors, namely :
\begin{equation}
\mathcal{F}(\gamma ^* N\to NP^0) =\chi_2^+
[
( \vec{\sigma }\cdot \vec{e}-\vec{e}\cdot \hat {\vec k} \ \vec{\sigma }\cdot 
\hat {\vec k})
f_t( k^2) +f_\ell ( k^2) \vec{e}\cdot \hat {\vec k}\ 
\vec{\sigma }\cdot \hat {\vec k} 
 ] \chi_1,
\end{equation}
where $\chi _1$ and $\chi _2$ are the two component spinors of the initial and
final nucleons\thinspace ; $f_t(k^2) $ and 
$f_\ell (k^2) $ are the threshold electromagnetic form factors, 
corresponding to the absorption of electric dipole
virtual photons with  transverse and longitudinal polarizations. 
At $k^2=0$, $f_\ell (0) =0$ and 
$f_t(0) = E_{0+}$ is the threshold electric dipole amplitude
for $\gamma +N\to N+\pi $ (with real photons).

In the framework of the {\it IA} (Fig. 3) the form factors 
$f_{1t}(k^2) $, $f_{1\ell }(k^2) $ and $f_2(k^2) $ for 
$\gamma ^{*}+d\to d+P^0$ can be directly related to the form factors 
$f_\ell (k^2) $ and $f_t(k^2) $ for $\gamma
^{*}+N\to N+P^0$.

\section{Impulse Approximation}

The most conventional starting point of possible mechanisms for pion 
electroproduction on the deuteron is the {\it IA} . This is, for 
example, the main mechanism in the region of $\Delta$-excitation, where the 
rescattering effects for $\gamma+d\to d+\pi^0$ are negligible 
\cite{Ko77,Bo78}. A special attention has to be devoted to the threshold region, 
for $\gamma (\gamma^*)+d\to d+\pi^0$, in particular for pion 
electroproduction in S-state where the rescattering effects may play an 
important 
role. Nevertheless, it is possible to show, in a model independent 
way, using only the Pauli principle, that the main rescattering 
contribution due to the following two step process: 
$\gamma +d\to p+p+\pi^-(\mbox{and}~ n+n+\pi^+)\to d+\pi^0$ vanishes, when the 
two 
nucleons in the $NN\pi$-intermediate state are on mass shell.
We plan to discuss this problem in a separate paper.

\subsection{Isospin structure of the $\gamma^{*}+\lowercase{d}\to d+P^0$ and 
$\gamma^{*}+\lowercase{d}\to \lowercase{p}+\lowercase{n}+P^0$ reactions}

As it is well known, isospin is not conserved in electromagnetic 
interactions of hadrons, but the hadron electromagnetic current
has definite transformation properties relative to isospin symmetry. In 
general, this current contains an isoscalar and an isovector 
components. The isotopic spin of deuteron is equal to zero, therefore 
the amplitude of the $\gamma ^{*}+d\to d+\pi ^0$ process is defined by 
the isovector part of the electromagnetic current only. On the other hand the 
amplitude of the $\gamma ^{*}+d\to d+\eta $ reaction is defined by its 
isoscalar part. 

If the amplitude of the $\gamma + N\to N+\eta $ reaction (with real photons) is
driven in the near threshold region by the $S_{11}(1535)$ contribution, which is 
dominated by the isovector part, then the amplitude of 
$F(\gamma d\to d\eta ) $ must  be small. However the first $\gamma d\to d\eta $ 
experiment found a very large cross section \cite{An69}. During seventeen years 
any attempt to resolve this contradiction in the framework of quark models and
multipole analyses of the $\gamma^* N\to\pi N$ reaction, taking into 
account effects like rescattering, were unsuccessful. A dedicated experiment 
\cite{Kr95_2} with a tagged photon beam, found that the main contribution to the 
$d(\gamma,\eta )X$ reaction is due, in fact, to the inelastic deuteron break-up  
$\gamma +d\to \eta +n+p$. In this process, the isovector nature of the 
transition $\gamma +N\to S_{11}(1535)$ results in the production of a 
$(np)$-system with isotopic spin $I=1$. Therefore, near the threshold 
of the $\gamma +d\to \eta+n+p$ reaction, it must be produced in a singlet state 
with $\mathcal{J}=0$. This simplifies drastically
the spin structure of the amplitude of the $\gamma +d\to d^*+\eta $~, 
$d^* =(n+p)_{\mathcal{J}=0}$ process since its coherent part must 
be determined essentially by the isovector ($i.e.$ large) 
part of the elementary 
$\gamma +N\to N+\eta $ process (in the framework of {\it IA}, (Fig. 4)).

In general, the amplitude for $\gamma +d\to n +p+P^0 $ (Fig. 5) contains an 
isoscalar and an isovector part :
\begin{equation}
F(\gamma d\to npP^0) =F_d(t)F(\gamma p\to pP^0)-F_d(u)
F(\gamma n\to nP^0),  
\label{eq:eq15}
\end{equation}
where $F_d$ is a generalized deuteron form factor, the variables $t$ 
and $u$ are the virtual $p$ and $n$ four-momentum squared. The minus sign 
in Eq. (\ref{eq:eq15}) is the consequence of the specific isotopic structure of the 
$d\to p^{*}+n$ and 
$d\to n^{*}+p$ vertices (with one virtual nucleon $N^{*}$).

At threshold the $u$ and $t$ variables are equal: 
$u_0=t_0=m^2-m_p ^2\displaystyle\frac m{2m+m_p}$ (in the 
limit: $M=2m$, $m$ is the nucleon mass ). Above threshold $u$ and $t$ are no 
longer equal. 

Rewriting Eq. (\ref{eq:eq15}):
\begin{eqnarray}
&F(\gamma d\to npP) =&\frac{1}{2}\left[ F_d(t) +F_d(u) \right]
\left[ F(\gamma p\to p P^0)-F(\gamma n \to nP^0) \right] \\
&& +\frac{1}{2}\left[ F_d(t) -F_d(u) \right] 
\left[ F(\gamma p \to p P^0) +F(\gamma n\to nP^0 ) \right], \nonumber
\end{eqnarray}
it is possible, by changing the variables $u$ and $t$ to control the 
relative role of the isoscalar and isovector contributions.

As mentioned above, the isotopic structure of the threshold amplitudes for
$\gamma +N\to N+\pi^0$  is a very actual problem. Both coherent 
processes, $\gamma +d\to d+\pi^0$ and $\gamma+d\to d^{*}+
\pi^0$, are sensitive to this structure but the  
$F(\gamma p\to p\pi^0)$ and $F(\gamma n\to n\pi^0)$ amplitudes contribute 
differently to these processes.
Therefore, the ratio of their cross sections near threshold will be essentially 
sensitive to the (small) $E_{0+}$ electric dipole absorption 
amplitude in $\gamma +n\to n+\pi ^0$. This ratio can be calculated 
using the existing experimental value for  $\gamma +p\to p+\pi ^0$ \cite{Fu96}:
$E_{0+}(\gamma p\to p\pi^0) =(-1,32\pm 0.05\pm 0,06)
\displaystyle\frac{e}{m_\pi}~10^{-3}$ and the theoretical predictions for 
$\gamma +n\to n+\pi^0$. For example, using the ChPT value as calculated in 
\cite{Ber96}:
$E_{0+}(\gamma n\to n\pi^0)=2.13 \displaystyle\frac{e}{m_\pi}10^{-3}$ one would 
get:
\begin{equation}
R=\displaystyle\frac{\sigma (\gamma d\to d^*\pi^0) }{\sigma(\gamma d\to 
d\pi^0)}=\displaystyle\frac{\left | S\right |^2}{\left|
V\right| ^2}=\frac{\left | 1.32+2.13\right |^2}{\left | 1.32-2.13\right|^2}
\simeq 18.
\end{equation}

If instead of ChPT predictions for $ \gamma + n\to n + \pi ^0$  we had taken 
dispersion relations calculations \cite{Ha97}, we would get  
$R\simeq 373$. We should notice that the dispersion relation calculation for 
neutron seems to be less stable than the one for proton. In any case these very 
large 
variations emphasize the large sensitivity of $R$ to the isotopic structure of 
the $\gamma + N\to N+\pi^0$ amplitudes.

Besides the real photon point, the $k^2$-dependence of $E_{0+}$ for 
both the  
$\gamma^*+p\to p+\pi ^0$ and $\gamma^* +n\to n+\pi ^0$ 
reactions is also very interesting \cite{Br97,Be94}.
 
\subsection{Relationship between the $\gamma ^{*}+d\to d+P^0$ and
$\gamma ^{*}+N\to N+P^0$ amplitudes}

In the framework of {\it IA}  (Fig. 3), the matrix element 
$\mathcal{M}( \gamma ^{*}d\to dP^0) $ for the  
$\gamma ^{*}+\;d\to d\;+\;P^0$ process can be written:
\begin{eqnarray}
\mathcal{M} &=&2\int d^3\vec{p}\mathcal{T}r\varphi ^{+}\left(
\left| \vec{p}+\frac 14\vec{Q}\right| \right) \hat{    
F}\varphi \left( \left| \vec{p}-\frac 14\vec{Q}\right|
\right),  \label{li26} \\
\vec{Q} &=&\vec k-\vec q,~~2\vec{p} =\vec{p_1}-\vec{p_2}+\frac 12    
\vec{Q},  \nonumber
\end{eqnarray}
where $\vec{P_1}=\vec{p_1}+\vec{p_2}$, 
$\vec{P_2}=\vec{p_1}^{^{\prime }}+\vec{p_2}$
, and $\vec k+\vec{p_1}=\vec q+    
\vec{p_1}^{^{\prime }}$ (the notation is explained in Fig. 3),
$$
F( \gamma N\to NP^0) =\chi _2^+\hat{F}\chi _1
, 
$$
\begin{equation}
\hat{F} =\left( \vec{\sigma }\cdot \vec K+L\right) /2
\end{equation}
and $\vec K$, $L$ are the spin-dependent and spin-independent
contributions to the matrix $\hat{F}$.

For the deuteron wave function we shall use the following representation,
which takes into account the $S$- and $D$-waves in the $np$-system :
\begin{eqnarray}
\varphi ( \vec{p}) &=&\frac 1{( 2\pi )
^{\frac 32}}\int d^3\vec{r} e ^{-i\vec{p}
\vec{r}}\varphi ( r),  \label{li28} \\
\varphi ( r) &=&\frac 1{\sqrt{4\pi }}\left[ \vec{    
\sigma }\cdot \vec{D}\frac{u(r)}{r}+\frac{w(r) }{\sqrt{2}r}
\left( 3\frac{\vec{\sigma }\cdot \vec{r    
}\vec{D}\cdot \vec{r}}{r^2}-\vec{\sigma }\cdot     
\vec{D}\right) \right] \frac{i\sigma _2}{\sqrt{2}}, 
\nonumber
\end{eqnarray}
where $u(r) $ and $w(r) $ are the standard
wave functions of the $S$- and $D$-states in deuteron. Expression (18) is 
particularly convenient to establish the spin structure of the amplitude of
the $\gamma ^{*}+d\to d+P^0$ process.
Since in general the amplitudes $\vec K$ and $L$ (for
the processes $\gamma ^{*}+N\to N+P^0$) depend on the integration
momentum $\vec{p}$ in (18), the wave functions $u $ and $w$ of the initial and 
final deuteron will not depend 
on the same variable. Indeed, due
to the nonlocality of $\gamma ^{*}N\to NP^0$ vertex, the coordinates 
$\vec{r}$ and $\vec{r}^{^{\prime }}$ of the initial and
final deuterons do not coincide. However, choosing the $\vec K$ and $L$ 
amplitudes at a particular
value of the internal momentum $\vec{p_1}$, $\hat{F}$ can be taken 
outside the integration symbol. This allows to express
the quantity $\mathcal{M}$  in terms of a definite combination of
deuteron form factors,  multiplied by the isovector (isoscalar) amplitudes
for the $\gamma ^{*}+N\to N+\pi ^0$ $( \gamma ^{*}+N\rightarrow
N+\eta ) $ reaction ({\it factorization hypothesis}).

This procedure is usually justified by a rapid fall-off of 
$\varphi(\left|\vec{p})\right|$ when $\left|\vec{p}\right|$ increases and by a 
(relatively) weak dependence of the $\vec K$ and $L$ amplitudes on 
$\left| \vec{p}\right|$.

After some transformations, Eq (18) becomes:
\begin{eqnarray}
\mathcal{M}(\gamma^*d\to dP^0) &=&\vec{D_1}\vec{D_2^*}\hat{L}F_1(\vec{Q^2})
+2\left (3\vec{D_1}\cdot \hat{\vec Q}\vec{D_2^*}
\cdot{\vec Q}-\vec{D_1}\cdot\vec{D_2}\right ) 
\hat{L}F_2(\vec{Q^2})\nonumber \\
&&
+i\hat{\vec K}\cdot \vec{D_1}\times\vec{D_2^*}\left(F_3(\vec{Q^2})
+F_4(\vec{Q^2})\right)-3i\hat{\vec K}\cdot \hat{\vec Q}\hat{\vec Q}\cdot 
\vec{D_1}\times\vec{D_2^*}F_4(\vec{Q^2}), \label{li29} 
\end{eqnarray}
with $\hat{\vec Q}=\left(\vec k-\vec{q}\right) /| \vec k-\vec q| $, where 
$\hat{\vec K}$ and $\hat{L}$ are the values of $\vec K$ and $L$
for a definite value of $\vec{p_1}$ (see below).

The generalized deuteron form factors 
$F_i(\vec{Q^2})$ are defined by :
$$F_1(\vec{Q^2}) ={\int }_0^{\infty}dr~j_0
\left(\displaystyle\frac{Qr}{2}\right)
\left [u^2(r)+w^2(r)\right ],$$
$$F_2(\vec{Q^2}) ={\int }_0^{\infty}dr~j_2\left(\frac{Qr}2\right) 
\left[u(r)-\frac{w( r)}{\sqrt{8}}\right] w(r),$$
\begin{equation}
F_3(\vec{Q^2}) ={\int }_0^{\infty}dr~j_0\left(\frac{Qr}2\right)
\left[u^2(r)-\displaystyle\frac{1}{2}w^2(r)\right], 
\end{equation}
$$F_4(\vec{Q^2})=\int_0^{\infty}dr~j_2
\left(\frac{Qr}2\right) \left[ u(r) +\frac 1{\sqrt{2}}w(r) \right]w(r),$$
$$j_0( x) =\frac{\sin x}x,\ j_2(x)=\sin x\left(\frac 3{x^3}-\frac 1x\right) 
-3\frac{\cos x}{x^2}.
$$
The combinations of the deuteron wave functions $u(r)$ and $w(r)$ in 
$F_i( \vec{Q^2})$ define the
charge, the magnetic and quadrupole form factors of the deuteron. 
The fourth form factor $F_4$ in Eq. (22) is associated with a
nonconservation of the current of the transition $d\to d+\pi^0$, due to 
the
specific structure of the triangle diagram contribution.

The calculated form factors, $F_i(\vec{Q^2})$, using Bonn \cite{Ma87} and Paris 
\cite{La80} deuteron wave functions, are shown in Fig. 6.

The quantity $\vec{Q^2}$ characterizes the value of the 
four-momentum transfer squared $t$ in the $\gamma ^{*}+d\to d+P^0$ 
reaction, \[
t=( k-q) ^2=2M\left( M-\sqrt{M^2+\vec{Q^2}}\right), 
\]
so that $t\equiv -\vec{Q^2}$, when $\left| \vec{Q}\right| \ll M$.

Obviously, the structure  of the $\gamma ^{*}+d\to d+P^0$ amplitude, 
Eq. (21), is not the most general one, even in the case of arbitrary values 
of $\vec K$ and $L$ and deuteron form factors $F_i(\vec{Q^2})$. Let us consider 
first the general spin structure of the
amplitude for $\gamma ^{*}+N\to N+P^0$ process\thinspace :
$${\cal M}(\gamma^*N\to N\pi)=\chi^{\dagger}_2{\cal F}\chi_1,$$
\begin{equation}
{\cal F}=i\vec e\cdot\hat{\vec k}\times\hat{\vec q} f_1+
\vec\sigma\cdot\vec e f_2 +\vec\sigma\cdot\hat{\vec k}~
\vec e\cdot\hat{\vec q}f_3+\vec\sigma\cdot\hat{\vec q}~\vec e\cdot\hat{\vec 
q}f_4
\end{equation}
$$+\vec e\cdot\hat{\vec k}(\vec\sigma\cdot\hat{\vec 
k}f_5+\vec\sigma\cdot\hat{\vec q}f_6),$$
where $f_i=f_i( s_1,t,k^2) $ are the scalar amplitudes for 
$\gamma^{\ast}+N\to N+P^0$, so that
$$
L =if_1\vec{e}\cdot \hat{\vec k}\times \hat{\vec q}, 
$$
\begin{equation}
\vec K  =\vec e \cdot f_2+
\hat{\vec k}\left( \vec e \cdot      
\hat{\vec q}f_3+\vec e \cdot \hat{\vec k}f_5\right) +   
\hat{\vec q}\left( \vec e \cdot \hat{\vec q}f_4+     
\vec {e}\cdot \hat{\vec k} f_6\right).  
\end{equation}
Comparing the expression (21) for the amplitude 
${\cal M}(\gamma^{\ast}d\to d P^0 )$ in {\it IA} with the general
spin structure of the amplitude one can establish a definite connection
between both sets of scalar amplitudes, namely $g_i$, $i=1-13$, for 
$\gamma^ * +d\to d+P^0$ (on one side) and $f_k$, $k=1-6$, for 
$\gamma^* +N\to N+P^0$ (on another side). Their exact relations are
given below:
\begin{eqnarray}
g_1 &=&-g_3=\sin\theta(f_3+\cos\theta f_4)\left( F_3\left( 
\vec{Q}^2\right)+F_4\left( 
\vec{Q}^2\right)\right) \nonumber  \\
&&-3Q_k\left( Q_mf_2+Q_k\sin\theta f_3+Q_q\sin\theta 
f_4\right)
F_4\left( \vec{Q}^2\right) ,\nonumber  \\
g_2 &=&-g_4=-(f_2+\sin^2\theta f_4)\left( F_3\left( \vec{Q}^2\right)+F_4\left( 
\vec{Q}^2\right)\right) \nonumber  \\
&&
+3Q_m\left( Q_mf_2+Q_k\sin\theta+Q_q\sin\theta  
f_4\right)
F_4\left( \vec{Q}^2\right) , \nonumber \\
g_5 &=&\sin\theta f_1\left[ F_1\left( \vec{Q}^2\right) +2F_2\left( 
\vec{Q}^2\right)\left(
3Q_m^2-1\right) \right] ,\nonumber \\
g_6 &=&\sin\theta f_1\left[ F_1\left( \vec{Q}^2\right) -2F_2\left( 
\vec{Q}^2\right) \right] , \nonumber\\
g_7 &=&\sin\theta f_1\left[ F_1\left( \vec{Q}^2\right) +2F_2\left( 
\vec{Q}^2\right)\left(
3Q_k^2-1\right) \right] , \nonumber\\
g_8 &=&6\sin\theta Q_mQ_kf_1F_2\left( \vec{Q}^2\right) -f_2\left(
F_3\left( \vec{Q}^2\right)+F_4\left( \vec{Q}^2\right)\right) , \nonumber\\
g_9 &=&6\sin\theta Q_mQ_kf_1F_2\left( \vec{Q}^2\right) +f_2\left(
F_3\left( \vec{Q}^2\right)+F_4\left( \vec{Q}^2\right)\right) ,\nonumber \\
g_{10} &=&-g_{12}=(f_2+f_3\cos\theta+f_4\cos^2\theta+f_5+f_6\cos\theta 
\left( F_3\left( \vec Q^2\right)+F_4\left( 
\vec Q^2\right)\right)  \nonumber \\
&&-3Q_k[ Q_k (f_2 +f_3\cos\theta + f_5 )+Q_q (\cos\theta f_4+f_6)  ]
F_4\left( \vec Q^2\right) , \nonumber \\
g_{11} &=&-g_{13}=-\sin\theta (cos\theta f_4+f_6)
 [ F_3\left( \vec {Q}^2\right)+F_4 
(\vec Q^2)] +3Q_m[ Q_k(f_2+\cos\theta f_3+f_5)
+\nonumber \\
&& Q_q(\cos\theta f_4+f_6)
F_4\left( \vec Q^2\right), \nonumber
\end{eqnarray}
where $Q_m=\hat{\vec Q}\cdot \hat {\vec {m}}$, ${Q}_k=     
\hat{\vec Q}\cdot \vec k$, $Q_m^2+Q_k^2=1$, $
Q_m^2=\sin ^2\theta \displaystyle\frac{\vec q^2}{| \vec k-     
\vec q| ^2}$, and $\theta $ is the $P^0$-meson production
angle in $CMS$ of $\gamma ^{*}+d\to d+P^0$ process.

Note that the relations $
g_1+g_3=g_2+g_4=g_{10}+g_{12}=g_{11}+g_{13}=0,$
which are correct for any amplitude $f_k$, result from the factorization 
hypothesis.

Neglecting the $D$-wave contribution, we can predict that the following ratios:
$$
\left( H^{(0)}_{xx}-H^{(0)}_{yy}\right) /\left( H^{(0)}_{xx}+H^{(0)}_{yy}\right)
_0 =\left(|f_3|^2+|f_4|^2-|f_1|^2-|f_2|^2\right )/ 
\left(|f_1|^2+|f_2|^2+|f_3|^2+|f_4|^2\right ),
$$
\begin{equation} 
\left( H^{(0)}_{zz}\right) /\left( H^{(0)}_{xx}+H^{(0)}_{yy}\right)
_0 =\left(|f_5|^2+|f_6|^2\right )/ \left(|f_1|^2+|f_2|^2+|f_3|^2+|f_4|^2\right 
),
\end{equation}
which do not depend on deuteron form factors and therefore on the deuteron 
structure.

\subsection{Threshold $\pi^0$ electroproduction within the impulse 
approximation}

At threshold the  $L$ and $\vec K$ amplitudes for $\gamma ^{*}+N\to 
N+P^0$ 
reduce to:
$$\vec K =f_t\left( \vec{e}-\hat{\vec k}\vec{e}\cdot \hat{\vec k}\right) +
f_\ell \hat{\vec k}\vec{e}\cdot \hat{\vec k},~~L =0,$$
where $f_\ell (k^2) $ and $f_t(k^2) $ are the
threshold form factors for $\gamma ^{*}+N\to N+\pi ^0$, corresponding to
absorption of electric dipole virtual photons with longitudinal and
transverse polarizations.

Taking into account the fact that, at  threshold, 
$\hat{\vec Q}= \hat{\vec k}$, one obtains the following form factors for 
$\gamma ^{*}+d\to d+\pi ^0$, which are
correct in the framework of the {\it IA} :
$$
f_{1t}(k^2) =f_t(k^2) \left( F_3+F_4\right),
~~f_{1\ell }(k^2) =f_\ell (k^2) \left(
F_3-2F_4\right) ,~~f_2(k^2) =0,
$$
$i.e.$ the magnetic quadrupole form factor $f_2(k^2) $ is equal
to zero in this approximation, independently of the deuteron structure.

\section{Model for $\gamma^*+N\to N+\pi^0$}

In order to calculate the scalar amplitudes $g_i,~i=1-13$, for the 
$\gamma^*+d\to d+\pi^0$ process in framework of {\it IA} , it is necessary to 
know the $\vec Q^2$-dependence of the deuteron form factors $F_j(Q^2),~j=1-4$, 
from one side, and the elementary amplitudes $f_k,~k=1-6$, for the process 
$\gamma^*+N\to N+\pi^0$, from another side.
In order to calculate the isovector part of the amplitudes $f_k$ for 
$\gamma^*+N\to N+\pi^0$, we shall use the effective Lagrangian approach-with a 
standard set of contributions (Fig. 7). Such model has successfully 
reproduced the experimental data  \cite{Fr99}, for the process $e+p\to 
e+p+\pi^0$ in the following kinematical conditions: $1.1\le W\le 1.4$ GeV and 
$-k^2=2.8$ and 4.0 (GeV/c)$^2$, in the whole domain of cos $\theta_{\pi}$ and 
azymuthal angle $\phi$. The main ingredients of this calculation were the s- and 
u-channel contributions of N and $\Delta$, with particular attention to the 
'off-shell' properties of the $\Delta-$isobar. The comparison with the 
experimental data allowed to determine the following values of the 
electromagnetic form factors for the $\gamma^*+p\to \Delta^+$ transition:
$$G_M^*/3G_d,~~G_d=(1-k^2/0.71~\mbox{GeV}^2)^{-2},~~R_{EM}=E_{1+}/M_{1+},~R_{SM}
=S_{1+}/M_{1+},$$
where $M_{1+}, ~E_{1+}$ and $S_{1+}$ denote the magnitude of the magnetic 
dipole, electric (transversal) quadrupole and Coulomb (longitudinal) quadrupole 
amplitudes or transition form factors for the $\gamma^*+N\to\Delta$ excitation.
In our analysis we will use the two following results of the experiment 
\cite{Fr99}:
\begin{itemize}
\item The magnetic dipole form factor $G_M^*(k^2)$ dominates, i.e. the ratios 
$R_{EM}$ and $R_{SM}$ are small (in absolute value);
\item The magnetic dipole form factor $G_M^*$ decreases with $-k^2$ faster than 
the dipole formula.
\end{itemize}

We parametrize the inelastic magnetic form factor  $G_M^*(k^2)\equiv G(k^2)$ for 
the $N\to \Delta$ electromagnetic transition with the help of  the following 
formula:
$$G(k^2)=\displaystyle\frac{G(0)G_d(k^2)}{(1-k^2/m_x^2)}.$$
Using the last experimental data about the ratio $G(k^2)/3G_d$ we find 
$m^2_x=5.75$ (GeV/c)$^2$, in agreement with previous estimates \cite{La75}.

Note in this connection, that the new JLab data \cite{Cfp99} about the electric 
proton form factor $G_{Ep}(k^2)$ show also a deviation from the dipole formula, 
with a similar value of the parameter $m_x$.

In order to calculate the amplitudes $f_i$, $i=1,..6$, for the elementary 
processes $e^-+N\to e^-+N+\pi^0$, $N=p$ or $n$, we will use a model similar to 
\cite{Va95} but with the following modifications:
\begin{itemize}
\item we introduce a term describing the exchange of $\omega$-meson in the 
$t-$channel;
\item the $s$-channel contribution of the $\Delta$ isobar is parametrized in 
such a form to avoid any off-mass shell effects (such as the admixture of 
$1/2^{\pm}$ or $3/2^-$ states).
\item the $u-$channel of the $\Delta-$isobar is neglected.
\end{itemize}

In order to justify the last option, 
let us note the essential difference between 
the $u-$channel contributions of $N$ and $\Delta$. The necessity to introduce 
the $u$-channel contribution from the proton exchange in the process 
$\gamma^*+p\to p+\pi^0$ is 
dictated by the gauge invariance of the electromagnetic interaction. As a 
byproduct, it derives the crossing symmetry for the resulting $s+u$ proton 
exchange. In case of $\Delta$-exchange, there is a different situation with 
respect to the above mentioned symmetry properties: the gauge invariance and the 
crossing symmetry. Due to the non-diagonality of the electromagnetic transition 
$\gamma^*+N\to \Delta$, it is possible to parametrize this vertex in a gauge 
invariant form independently from the virtuality of the $\Delta$. Therefore, the 
$\Delta$-contribution only in the $s$-channel, is gauge invariant, 
independently from the $u$-channel $\Delta$-contribution. This means that for 
the $\Delta$-contribution there is no direct connection between the gauge 
invariance and the crossing symmetry, as for the proton exchange. Moreover, even 
the $\Delta$-contribution in $s-$ and $u$- channels simultaneously will not 
induce crossing symmetry. Namely due to the presence of the $\Delta$-pole in the 
physical region of s-channel, it is necessary to introduce the $\Delta-$width in 
the corresponding propagator- with resulting complex amplitudes, whereas the 
$u-$channel $\Delta$-contribution is characterized by real amplitudes. It turns 
out that  we do not have exact crossing symmetry for the $\Delta$-contributions, 
even for the sum of $u$-and $s$ diagrams with $\Delta$-exchange. 

We will consider only the $s-$ channel $\Delta$-contribution. In order to avoid 
problems with off-mass shell effects, we write the matrix element for the 
$\Delta$-contribution  following the two-component formalism for the description 
of the spin structure of both vertices, $\Delta \to N+\pi$ and $\gamma^*+N\to 
\Delta$. Therefore we can write:
$$\gamma+N\to \Delta:~~\vec e\times\vec 
k\cdot\vec\chi^\dagger{\cal I}\chi_1 ,~~\mbox{M1~transition,~only!}$$
$$\Delta\to N+\pi:~~\chi_2^\dagger{\cal I}\vec\chi\cdot\vec q,$$
where ${\cal I}$ is the identity matrix. Each component of the vector $\vec\chi$ 
is a 2-component spinor, 
satisfying the condition $\vec\sigma\cdot\vec\chi=0$, in order to avoid any spin 
1/2 contribution. Using for the $\Delta$ density matrix the following 
expression:
$$\rho_{ab}=\displaystyle\frac{2}{3}(\delta_{ab}-
\displaystyle\frac{i}{2}\epsilon_{abc}\sigma_c),
$$
we can write the matrix element for the $\Delta$-contribution in the CMS of 
$\gamma^*+N\to N+\pi^0$ as follows:
\begin{equation}
{\cal M}_{\Delta}=
\displaystyle\frac{eG(k^2)|\vec q|}
{M^2_{\Delta}-s-i\Gamma_{\Delta}M_{\Delta}}\chi_2^{\dagger}
(2i\vec e\cdot\hat{\vec k}\times \hat{\vec q}+cos\theta_{\pi}
\vec\sigma\cdot\vec e-\vec\sigma\cdot\hat{\vec k}\vec e\cdot\hat
{\vec q})\chi_1 \sqrt{(E_1+m)(E_2+m)},
\end{equation}
where
$M_{\Delta}~(\Gamma_{\Delta})$ is the mass (width) of $\Delta$.

The following $\Delta$ contributions to the scalar amplitudes, 
$f_{i\Delta},~i=1-6$, can be derived:
$$ f_{1\Delta}=2\Pi(s,k^2),$$
$$ f_{2\Delta}=\cos \theta_{\pi}\Pi(s,k^2),$$
\begin{equation} f_{3\Delta}=-\Pi(s,k^2),\end{equation}
$$ f_{4\Delta}=f_{5\Delta}=f_{6\Delta}=0,$$
where we use the notation:
$$\Pi(s,k^2)=\displaystyle\frac{G(k^2)|\vec q|}
{M^2_\Delta -s-i\Gamma_\Delta M_\Delta}.$$
The normalization constant $G(0)$ can be deduced from the value of the total 
cross section for the reaction $\gamma+p\to p+\pi^0$ (with real photons) at 
$s=M_{\Delta}^2$:
$$\sigma_T(\gamma p\to p\pi^0)=\displaystyle\frac{\alpha}{2}G^2(0)
\displaystyle\frac{|\vec q|^3}
{|\vec k|}\displaystyle\frac{(E_{1\Delta}+m)(E_{2\Delta}+m)}{M_{\Delta}^4
\Gamma^2_{\Delta}},$$
where 
$$E_{1\Delta}=\displaystyle\frac{M^2_{\Delta}+m^2}{2M_{\Delta}},
~~E_{2\Delta}=\displaystyle\frac{M^2_{\Delta}+m^2-m^2_{\pi}}{2M_{\Delta}},$$
$$|\vec k|=\displaystyle\frac{M^2_{\Delta}-m^2}{2M_{\Delta}},~~
|\vec q|=\sqrt{E_{2\Delta}^2-m^2}.$$
Using the spin structure of the resonance amplitude (27), we obtain the
following structure for the resonance contribution to the matrix element of the 
process $\gamma^*+d\to d+\pi^0$:
$$
{\mathcal{M}}_\Delta(\gamma^{*}d\to d\pi )=\displaystyle\frac{1}{2}\Pi(s,k^2) 
\left \{ 2\sin \theta \vec{e}\cdot\hat {\vec n}
\left [ F_1\left(\vec {Q^2}\right )\vec{D_1}\cdot \vec{D_2}^{*} 
+F_2\left(\vec {Q^2}\right )
( 3\vec{D_1}\cdot \hat{\vec Q} \vec{D_2}^*\cdot \hat{\vec Q}-\vec{D_1}\cdot 
\vec{D_2}^*) \right ]\right .$$
$$ +\left [\left ( \vec{e}\cdot \hat {\vec {m}}~\hat {\vec {m}}\cdot      
\vec{D_1}\times \vec{D_2}^*+\vec{e}\cdot      
\hat {\vec n}~\hat {\vec n}\cdot\vec{D_1}\times 
\vec{D_2}^*\right ) \cos\theta -\vec{e}\cdot \hat {\vec q}~\hat {\vec k}\cdot      
\vec{D_1}\times \vec{D_2}^{*} \right ]
\left ( F_3\left (\vec {Q^2}\right )+F_4\left(\vec {Q^2}\right )\right )$$
$$  
\left . -3 \cos\theta F_4\left(\vec {Q^2}\right )\vec{e}\cdot \hat {\vec {m}}
\hat{\vec {Q}}\cdot \vec{D_1}\times 
\vec{D_2}^* Q_m + 3 F_4\left(\vec {Q^2}\right )\vec{e}\cdot\hat {\vec q} \vec 
Q\cdot  \vec{D_1}\times 
\vec{D_2}^* Q_k\right \}.$$
Taking into account only the S-wave component of the deuteron wave function
it is possible to  predict the $\theta-$ dependence for the simplest
polarization observables for $\gamma ^{*}+d\to d+\pi ^0$:
$$\left( H^{(0)}_{xx}-H^{(0)}_{yy}\right) /\left( H^{(0)}_{xx}+
H^{(0)}_{yy}\right) =-3\frac{\sin^2\theta }{3-2\cos ^2\theta },~~
H^{(0)}_{xz} =H^{(0)}_{zz}=0.
$$
and in the case of tensor polarized deuterons :
\begin{eqnarray*}
\left( H^{(2)}_{xx}+H^{(2)}_{yy}\right) /\left( H^{(0)}_{xx}+H^{(0)}_{yy}\right)
_0 &=&-Q_{zz}\frac{\cos ^2\theta }{4\left( 3-2\cos ^2\theta \right) },
\\
\left( H^{(2)}_{xx}-H^{(2)}_{yy}\right) /\left( H^{(0)}_{xx}+H^{(0)}_{yy}\right)
_0 &=&\left( Q_{xx}-Q_{yy}\right) \frac{\cos ^2\theta }{4\left( 3-2\cos
^2\theta \right) }\text{.}
\end{eqnarray*}

For comparison, note that in the case of the process $e+p\to e+p+\pi 
^0$ we have 
(for an unpolarized proton target):
\[
\left( H^{(0)}_{xx}-H^{(0)}_{yy}\right) /\left( H^{(0)}_{xx}+
H^{(0)}_{yy}\right) =-\frac{5\sin^2\theta }{5-3\cos ^2\theta }. 
\]

The matrix element ${\cal M_{\omega}}$ for the $\omega$-exchange in 
$\gamma^*+N\to N+\pi^0$ can be written in the following form:

$${\cal M}_{\omega}=\displaystyle\frac{g_{\omega}G_{\omega}(k^2)}
{m_{\omega}(t-m_\omega^2)}\epsilon_
{\mu\nu\rho\sigma} e_{\mu}k_{\nu}q_{\sigma}u(p_2)
\left[\gamma_{\rho}-\displaystyle\frac{\kappa_{\omega}}{2m}
\sigma_{\rho\beta}(k-q)_\beta \right ]u(p_1)$$
The constants $\kappa_{\omega}$ and  $g_{\omega}$ are fixed by the Bonn 
potential 
\cite{Ma87}: $\kappa_{\omega}=0,$ $g_{\omega}^2/4\pi=20.$ The VDM suggests the 
following parametrization for the form factor $G_{\omega}(k^2)$:
$$G_{\omega}(k^2)=\displaystyle\frac{G_{\omega}(0)}{1-k^2/m^2_{\rho}}.$$
The value $G_{\omega}(0)$ can be fixed by the width  of the radiative decay 
$\omega\to \pi\gamma$, through the following formula:
$$\Gamma(\omega\to\pi\gamma)=
\displaystyle\frac{\alpha}{24}G_{\omega}^2(0)
\left ( 1-\displaystyle\frac{m_{\pi}^2}{m_{\omega}^2} \right )^3 m_{\omega},$$
where 
$BR(\omega\to \pi^0\gamma)=
\Gamma(\omega\to \pi^0\gamma)/\Gamma_{\omega}=(8.5\pm 1.5)\%$, 
$\Gamma_{\omega}=(8.81\pm 0.09)$ MeV and $m_{\omega}=782$ MeV.

Concerning vector meson exchange in $e^-+N\to e^-+N+\pi$, it is known 
\cite{Va95}, that the vector meson exchange is 
important for the processes $\gamma+N\to N+\pi$, in the considered region of 
$W$. Due to the isovector nature of the electromagnetic current in 
$\gamma^*+d\to d+\pi^0$, 
the $\rho^0$-contribution to $\gamma^*+N\to N+\pi^0$ is exactly 
cancelled. The VDM parametrization of the electromagnetic form factors 
suggested above for the $\gamma^*\pi\omega$-vertex as to be considered as a 
simplified possibility for the space-like region of momentum transfer, where 
there is no experimental information. However, in the region of time-like 
momentum transfer, different pieces of information exist. Let us mention three 
of them. The decay $\omega\to \pi+\ell^++\ell^-$ \cite{pdgom} allows to measure 
this 
form factor in the following region $4m_\ell\le k^2\le (m_\omega-m_\pi)^2,$ 
where $m_{\ell}$ is the lepton mass. The process $e^+ +e^-\to \pi^0+\omega$ 
\cite{pdgem} is driven by the considered form factor in another time-like 
region, 
namely for $k^2\ge(m_\omega+m_\pi)^2$. For completeness we mention the 
$\tau^-\to \nu_{\tau}+ \pi^- +\omega$ decay \cite{pdgtau}. The presence of the 
same 
factor 
$G_{\omega}(k^2)$ in  processes so different as $e^+ +e^-\to \pi^0+\omega$ and 
$\tau \to \nu_{\tau}+\pi^-+\omega $ results from the well known 
CVC hypothesis (Conservation of Vector Current for the weak semileptonic 
processes).

Note also that we have taken a 'hard' expression for the VDM form factor, 
$G_{\omega}(k^2)$,  which  is assumed to reproduce at best the structure 
function $A(k^2)$  of elastic $ed$ scattering, through the calculations of the 
meson exchange current due to $\pi\rho$ exchange\cite{VO95}.
However this conclusion is correlated to the properties of the nucleon form 
factor, especially with the behavior of the isoscalar electric form factor, 
$G_{Es}=(G_{Ep}+G_{En})/2.$ New $G_{Ep}$ data \cite{Cfp99} 
(with large deviation from the previously assumed dipole behavior) will also 
favor a hard form factor 
$G_{\omega}(k^2)$ for the good description of the $k^2$ dependence of $A(k^2)$ 
at large momentum transfer. However a satisfactory description will depend also 
on the large $k^2$-dependence of the neutron electric form factors, which will 
be 
measured in the next future up to $\left |k^2\right |=2 $ (GeV/c)$^2$ 
\cite{Madey}.
It is then expected that the different observables in the processes $e+N\to 
e+N+\pi^0$ and $e+d\to e+d+\pi^0$ at relatively large momentum transfer are 
sensitive to the parametrizations of the form factor $G_{\omega}(k^2)$. 
For example, the VDM parametrization for $G_{\omega}(k^2)$ shows that this form 
factor is 
'harder' in comparison with nucleon and $N\to \Delta$ form factors. Therefore,
in this case, the relative role of $\omega$-exchange  will be essentially 
increased at large momentum transfer.

\section{Results and discussion}

In order to test the model for $\pi^0$ electroproduction on deuterons, we 
compared our calculation to experimental data on  $\pi^0$ and $\pi^+$  
photoproduction on proton in the $\Delta$-resonance region. The 
angular distributions at different energies of the real photon reproduce quite 
well the existing data, a sample of which is shown in Fig. 8. This agreement 
justifies the generalization of the model in case of 
$\pi^0$-electroproduction on nucleons, $e^-+N\to e^-+N+\pi^0$, by introducing 
the corresponding electromagnetic form factors in the different photon-hadron 
vertices (see Fig. 7). Note also, that the resulting electromagnetic 
current for the process $\gamma^*+N\to N+\pi^0$ (with virtual photon) still 
satisfies the gauge invariance, for any parametrization of the electromagnetic 
form factors, and for any values of the kinematical variables $k^2$, $W$ and 
$\cos\theta_{\pi}$.
However this model does not satisfy the T-invariance of the electromagnetic 
interaction, but here we will consider only T-even observables, such as the 
different contributions to the $d(e,e\pi^0)d$ differential cross section ( with 
unpolarized particles in the initial and final states). This problem, which is 
common to all modern approaches of pion photo- and electro-production on 
nucleons, is generally not discussed in the existing literature.

In the framework of {\it IA} , as it was shown before, the deuteron structure is 
described by by four inelastic form factors $F_i(\vec Q^2),~i=1-4$, where the 
argument $\vec{Q^2}$ depends on all the three kinematical variables, $k^2$, $W$ 
and 
$\cos\theta_{\pi}$, which characterize the process $\gamma^*+N\to N+\pi$:
$$\vec Q^2=(\vec k-\vec q)^2=\vec k^2+\vec {q^2}-2|\vec k||\vec q|\cos 
\theta_{\pi},$$
with
$$\vec k^2=k_0^2-k^2,~~k_0=\displaystyle\frac{W^2+k^2-m^2}{2W},$$
$$\vec q^2=E_{\pi}^2-m_{\pi}^2,
~~E_{\pi}=\displaystyle\frac{W^2+m_{\pi}^2-m^2}{2W}.$$
Fig. 9 illustrates the dependence of the variable $\vec Q^2$ on 
$\cos\theta_{\pi}$ at 
fixed values of $k^2$ and $W$, at W=1.2 GeV and W=1.137 GeV (which corresponds 
to $E_{\gamma}=220$ MeV, see Fig. 8). This dependence is similar for all values 
of $k^2$, in the interval $\left |k^2\right |=0.5\div 2.0 $ (GeV/c)$^2$. 
Note that $\vec Q^2_{max}\simeq 3$ (GeV/c)$^2$ at $-k^2=2$ (GeV/c)$^2$ , so, at 
the same 
value of four momentum transfer, the process $\gamma^*+d\to d+\pi^0$ is driven 
by the deuteron form factors at higher momentum transfer in comparison with 
elastic $ed$-scattering.

Comparing Fig. 9 and Fig. 6 (which shows the $\vec 
Q^2$-dependence of the deuteron form factors, in the 
interval 
$0\le \vec {Q^2}\le 3$ (GeV/c)$^2$), one can see that in the 
range $-k^2=0.5\div 2.0 $ (GeV/c)$^2$, the deuteron form factors are  very 
sensitive 
to the behavior of the deuteron wave function calculated in different 
NN-potentials.

The $\theta_\pi$-dependence of all four contributions to the inclusive 
$d(e,e\pi^0)d$ cross section, namely $H_{xx}\pm H_{yy}$, $H_{zz}$ and 
$H_{xz}+H_{zx}$, for  different values of $k^2$ and $W$ is shown in Figs. 10 and 
11 \footnote{Note that in our normalization, Eq. (2), all components 
$H_{ab}$ are dimensionless numbers.}. In order to show the relative role of the 
different mechanisms for the 
elementary processes $\gamma^*+N\to N+\pi$ ( in the considered kinematical 
region for the variables $k^2$ and $W$), each picture shows four curves: 
$\Delta$ contribution only, $\Delta+s+u$ (nucleon diagrams) and $\Delta+s+u\pm 
\omega $. The calculations are shown for both relative signs of the vector meson 
contribution in order to stress the importance of the $\omega$ contribution. The 
positive sign has been choosen from the comparison with experimental data on 
$\gamma+p\to p+\pi^0$ (real photons).

The $\omega$ contribution is important for all the four considered observables, 
in particular for the $H_{xx}\pm H_{yy}$ terms at $\theta_\pi\simeq 80^o$; in 
the case of  $H_{zz}$ the largest sensitivity appears for backward 
$\pi^0$ electroproduction.

The relative role of the absorption of virtual photon with longitudinal and 
transversal polarizations depends essentially on the variables $k^2$ and $W$, 
with an increase of the ratio $H_{zz}/(H_{xx}+ H_{yy})$ with $-k^2$. At 
W=1.2 GeV, where the $\Delta$-contribution (with absorption of transversal 
virtual photons) dominates, the relative role of $H_{zz}$ is weaker in 
comparison with $H_{xx}+ H_{yy}$. However for $-k^2\ge 1$ (GeV/c)$^2$ $ H_{zz}$ 
exceeds $H_{xx}+ H_{yy}$, even in the resonance region.

The ratio $(H_{xx}- H_{yy})/(H_{xx}+ H_{yy})$ is negative (due to the dominance 
of the transversal $\Delta$ and $\omega$-contributions) and has a $\simeq 
sin^2\theta_{\pi}$ 
behavior. The longitudinal-transversal interference contribution, 
$H_{xz}+ H_{zx}$, shows a particular sensitivity to the different ingredients of 
the model, with strong $\theta_{\pi}$-dependence, in the whole considered 
kinematical domain.

In view of the importance of the $\omega$ contribution to all observables for 
the $d(e,e\pi^0)d$ process, we studied the sensitivity to the choice of the 
electromagnetic $\gamma^*\omega\pi$-vertex form factor. For this aim we used two 
parametrizations, a $hard$ monopole form, $G_{\omega}^{(h)}(k^2)$, predicted by 
the standard VDM, and a 
$soft$ dipole form $G_{\omega}^{(s)}(k^2)$:
$$ G_{\omega}^{(h)}(k^2)=
\displaystyle\frac
{G_{\omega}(0)}
{ 1-\displaystyle\frac{k^2}{m^2_{\rho}}}, ~~
G_{\omega}^{(s)}(k^2)=
\displaystyle\frac
{G_{\omega}(0)}
{\left ( 1-\displaystyle\frac{k^2}{m^2_{\rho}}\right )^2}.$$
Fig. 12  shows the $\theta_\pi$-dependence of the following ratios:
$$r_{\pm}(\cos\theta_\pi)=
\displaystyle\frac
{(H_{xx}\pm H_{yy})_{hard}-(H_{xx}\pm H_{yy})_{soft}}
{(H_{xx}\pm H_{yy})_{hard}+(H_{xx}\pm H_{yy})_{soft}}$$
for two different values of $k^2$ ($-k^2=0.5$ and 2 (GeV/c)$^2$) and $W=1.137$. 
For $W=1.2$ GeV (Fig. 13) the largest sensitivity to 
the choice of the form factor $G_{\omega}(k^2)$ appears at forward angles for 
$\pi^0$-production, whereas at $W=1.137$ GeV all angles are equally sensitive to 
this choice.  At the $\Delta$-resonance this sensitivity increases slightly with 
$-k^2$. 

The absolute measurements of the different contributions to the inclusive cross 
section for $d(e,e\pi^0)d$ will help in defining
the appropriate $k^2$-dependence of the form factor $G_{\omega}(k^2)$.
However , as we can see on Fig. 14, the absolute values of the 
$H_{xx}\pm H_{yy}$ contributions, the shape and absolute values of 
$H_{zz}$ and  $H_{xz}+ H_{zx}$ are also sensitive to the existing 
$NN-$potentials, in particular at large $k^2$. In Figs 15, 16, 17 and 18, we 
illustrate the behavior of the four observables, for different parametrizations 
of the following ingredients:
\begin{itemize}
\item the deuteron wave function: for the Bonn \cite{Ma87} and Paris \cite{La80} 
potentials,
\item the electromagnetic form factors for the $\gamma^*\pi\omega$-vertex: 
$hard$ (VDM) and $soft$ (dipole) parametrizations;
\item the electromagnetic form factor of the proton: dipole or a 'softer' 
parametrization based on recent data on the proton electric form factor.
\end{itemize} 
The differences between the different parametrizations increase at large 
momentum transfer.

The inclusive cross section for $d(e,e)\pi^0d$ is characterized by two 
contributions, only. After integration over $d\Omega_{\pi}$, we have: 
$$ H_t(k^2,W)=\int_{-1}^{+1} d\cos\theta_{\pi}(H_{xx}+H_{yy}),$$
$$ H_\ell(k^2,W)=\int_{-1}^{+1} d\cos\theta_{\pi} H_{zz}.$$
The three-dimensional plot of Fig. 19 shows the dependence of these inclusive 
functions, on $k^2$ and $W$. The calculation is done here, for the $hard$ form 
factor $G_{\omega}$, the dipole form factor $G_{Ep}$ and the Bonn 
deuteron wave function.

\section{Conclusions}

We have made a general analysis of coherent pseudoscalar neutral mesons 
production on deuterons, $e+d\to e+d+P^0$, which holds for 
any kinematics of the discussed processes. Threshold 
$P^0$-meson production (at any value of momentum transfer square 
$k^2$ and for the minimum value of the effective mass of the produced
hadronic system) is especially interesting  due to the essential simplification
of the spin structure of the corresponding amplitudes and  to the decreasing 
number of independent kinematical variables. Another kinematical region, which 
is interesting for the process $\gamma ^{*}+d\to d+\pi ^0$, is the $\Delta 
$-isobar excitation on the nucleons. 

Coherent $P^0$-meson production is interesting due to its special 
sensitivity to the isotopic structure of the threshold amplitude for the 
elementary processes $\gamma^*+N\to N+P^0$.

The $\pi^0$-meson electroproduction on the deuteron allows to measure the 
threshold amplitude for $\gamma^* +n\to n+\pi^0$, which is 
important for testing hadron electrodynamics \cite{Ec95}.

The $\eta $-meson electroproduction on the deuteron could be important
for the study of $\eta N$- and $\eta d$-interactions, in particular after
the finding  of a strong energy dependence of the cross section of 
$n+p\to d+\eta$ process near threshold. 

The {\it IA} can be considered as a good starting point for the  discussion of  
corrections such as mesonic exchange currents, isobar configurations in 
deuteron, quark degrees of freedom, etc., but rescattering effects 
will also have to be discussed, in particular for 
$\eta$-production near threshold.

Using an adequate model for the elementary processes of 
$\pi^0$-electroproduction on nucleons, $e^-+N\to e^-+N+\pi^0$, which 
satisfactorily reproduces the angular dependence of the differential cross 
section for the processes $\gamma+p\to p+\pi^0$ and $\gamma+p\to n+\pi^+$ (in 
the $\Delta$-resonance region), we estimated the four standard contributions to 
the exclusive differential cross section for the reaction $d(e,e\pi^0)d$ as a 
function of the variables $k^2,W$ and $\theta_{\pi}$. These calculations were 
done at relatively large momentum transfer square, $-k^2=0.5\div 2.0$ 
(GeV/c)$^2$, 
where recent data exist. All observables show a large sensitivity to the 
parametrization of electromagnetic form factors, in the considered model. A 
special attention was devoted to the study of the effects of soft and hard 
parametrizations of form factor for the $\pi\omega\gamma^*$-vertex, as well as 
to possible deviation of the proton electric form factor from the dipole fit. 
Moreover, as it is well known for elastic $ed$-scattering, we find here, too, a 
large dependence of all the observables to the choice of $NN-$potential. The 
large sensitivity of the $d(e,e\pi^0)d$ cross section to the $\omega$-exchange 
contribution can be used, in principle, to study the corresponding 
electromagnetic form factors in the space-like momentum transfer region.

\noindent {\bf Acknowledgments}
\vspace{.5 true cm}

We thank J.-M. Laget for interesting 
discussions on rescattering effects.
One of the authors  (M. P. R.) is very indebted to the hospitality of 
Saturne where part of this work was done.

\begin{center}
\textbf{Appendix}
\end{center}
We present here the expressions for the structure functions $h_1-h_{41}$ in 
terms of the scalar amplitudes $g_1-g_{13}$. The SF's $h_1-h_5$ corresponding  
to the interaction with an unpolarized deuteron target can be written as:
\begin{eqnarray*}
3h_1 &=&\left| g_1\right| ^2+\left| g_2\right| ^2+\left| g_3\right| ^2+
\left| g_4\right| ^2, \\
3h_2 &=&\left| g_5\right| ^2+\left| g_6\right| ^2+\left| g_7\right| ^2+
\left| g_8\right| ^2+\left| g_9\right| ^2, \\
3h_3 &=&\left| g_{10}\right| ^2+\left| g_{11}\right| ^2+
\left| g_{12}\right| ^2+\left| g_{13}\right| ^2, \\
3h_4 &=&\cal{R}e~\left( g_1g_{10}^{*}+g_2g_{11}^{*}+g_3g_{12}^{*}+
g_4g_{13}^{*}\right), \\
3h_5 &=&\cal{I}m~\left( g_1g_{10}^{*}+g_2g_{11}^{*}+g_3g_{12}^{*}+
g_4g_{13}^{*}\right), \\
\end{eqnarray*}
We derive fhe following expressions for the $SF^{^{\prime }}s\ h_6$ - $
h_{18}$, which characterize the effects of the target vector polarization :
\begin{eqnarray*}
h_6 &=&-\cal{I}m~\left( g_2g_6^{*}-g_3g_9^{*}-
g_4g_7^{*}\right), \\
h_7 &=&\cal{I}m~\left(g_6g_{11}^{*}+g_7g_{13}^{*}+
g_9g_{12}^{*}\right), \\
h_8 &=&\cal{R}e~\left(g_2g_6^{*}-g_3g_9^{*}-
g_4g_7^{*}\right), \\
h_9 &=&\cal{R}e~\left(g_6g_{11}^{*}-g_7g_{13}^{*}-
g_9g_{12}^{*}\right), \\
h_{10} &=&-2\cal{I}m~g_1g_2^{*}, \\
h_{11} &=&-2\cal{I}m~\left( g_5g_9^{*}-g_7g_8^{*}\right), \\
h_{12} &=&-22\cal{I}m~g_{10}g_{11}^{*}, \\
h_{13} &=&-\cal{I}m~\left( g_1g_{11}^{*}-g_2g_{10}^{*}\right), \\
h_{14} &=&\cal{R}e~\left( g_1g_{11}^{*}-
g_2g_{10}^{*}\right), \\
h_{15} &=&\cal{I}m~\left( g_1g_6^{*}-g_3g_5^{*}-
g_4g_8^{*}\right), \\
h_{16} &=&\cal{I}m~\left( g_5g_{12}^{*}-g_6g_{10}^{*}-
g_8g_{13}^{*}\right), \\
h_{17} &=&-\cal{R}e~\left( g_1g_6^{*}-g_3g_5^{*}-
g_4g_8^{*}\right), \\
h_{18} &=&\cal{R}e~\left( g_5g_{12}^{*}-g_6g_{10}^{*}+
g_8g_{13}^{*}\right), \\
\end{eqnarray*}
Finally for the $SF^{^{\prime }}s\ h_{19}$ - $h_{41}$, which describe
the effects on tensor target polarization, one obtains :
\begin{eqnarray*}
3h_{19} &=&-\left| g_1\right| ^2+\left| g_2\right| ^2+\left| g_7\right| ^2-
\left| g_8\right| ^2, \\
3h_{20} &=&-\left| g_5\right| ^2+\left| g_9\right| ^2, \\
3h_{21} &=&-\left| g_{10}\right| ^2+\left| g_{11}\right| ^2, \\
3h_{22} &=&-\cal{R}e~\left( g_1g_{10}^{*}-g_2g_{11}^{*}\right), \\
3h_{23} &=&-\cal{I}m~\left( g_1g_{10}^{*}-g_2g_{11}^{*}\right), \\
3h_{24} &=&\left| g_2\right| ^2-\left| g_3\right| ^2-
\left| g_4\right| ^2, \\
3h_{25} &=&\left| g_6\right| ^2+\left| g_7\right| ^2+
\left| g_9\right| ^2, \\
3h_{26} &=&\left| g_{11}\right| ^2-\left| g_{12}\right| ^2-
\left| g_{13}\right| ^2, \\
3h_{27} &=&\cal{R}e~\left( g_2g_{11}^{*}-g_3g_{12}^{*}-
g_4g_{13}^{*}\right), \\
3h_{28} &=&\cal{I}m~\left( g_2g_{11}^{*}-g_3g_{12}^{*}-
g_4g_{13}^{*}\right), \\
3h_{29} &=&-2\cal{R}e~g_1g_2^{*}, \\
3h_{30} &=&-2\cal{R}e~\left( g_5g_9^{*}+g_7g_8^{*}\right), \\
3h_{31} &=&-2\cal{R}e~g_{10}g_{11}^{*}, \\
3h_{32} &=&-\cal{R}e~\left( g_1g_{11}^{*}+g_2g_{10}^{*}\right), \\
3h_{33} &=&\cal{I}m~\left( g_1g_{11}^{*}+g_2g_{10}^{*}\right), \\
3h_{34} &=&-\cal{R}e~\left( g_1g_6^{*}+g_3g_5^{*}+
g_4g_8^{*}\right), \\
3h_{35} &=&-\cal{R}e~\left( g_5g_{12}^{*}+g_6g_{10}^{*}+
g_8g_{13}^{*}\right), \\
3h_{36} &=&-\cal{I}m~\left( g_1g_6^{*}+g_3g_5^{*}+
g_4g_8^{*}\right), \\
3h_{37} &=&\cal{I}m~\left( g_5g_{12}^{*}+g_6g_{10}^{*}+
g_8g_{13}^{*}\right), \\
3h_{38} &=&-\cal{R}e~\left( g_2g_6^{*}+g_3g_9^{*}+
g_4g_7^{*}\right), \\
3h_{39} &=&-\cal{R}e~\left( g_6g_{11}^{*}+g_7g_{13}^{*}+
g_9g_{12}^{*}\right), \\
3h_{40} &=&-\cal{I}m~\left( g_2g_6^{*}+g_3g_9^{*}+
g_4g_7^{*}\right), \\
3h_{41} &=&\cal{I}m~\left( g_6g_{11}^{*}+g_9g_{12}^{*}+
g_7g_{13}^{*}\right).
\end{eqnarray*}


\vspace*{.5 true cm}
%
%
\begin{figure}
\caption{ One-photon exchange mechanism for the process $e+d\to e+d+P^0$.}
\end{figure}
\begin{figure}
\caption{ $np$-intermediate state contribution to the unitarity condition for 
$\gamma+d\to d+\eta$; the dotted line crosses the particles on mass 
shell.}
\end{figure}

\begin{figure}
 \caption{ {\it IA}  diagrams for  $\gamma+ d\to d+P^0$.}
\end{figure}
\begin{figure}
\caption{ {\it IA}  diagrams for the coherent part of the 
$\gamma+d\to p+n+\eta=\gamma+d\to d^*+\eta$ process.}
\end{figure}
\begin{figure}
\caption{  {\it IA}  diagrams for the incoherent part of the 
$\gamma+d\to p+n+\eta=\gamma+d\to d^*+\eta$ process.}
\end{figure}
\begin{figure}
\caption{$\vec{Q^2}$-dependence of the deuteron form factors, (see Eq. (22) )
$F_1$ (full line),  $F_2$ (dashed line),  $F_3$ (dotted line),  $F_4$ 
(dashed-dotted line). The calculation is  based on: (a)- the Paris wave 
function; (b) - the Bonn wave function.}
\end{figure}
\begin{figure}
\caption{ The Feynman diagrams for $\gamma^*+N\rightarrow N+\pi$- 
processes}
\end{figure}

\begin{figure}
\caption{The angular dependence of the differential cross sections for the 
photoproduction processes: (a) and (b) - $\gamma^*+p\rightarrow p+\pi^0$ full 
stars (open crosses) are data from \protect\cite{FH72} (\protect\cite{GH74}); 
(c) - $\gamma^*+p\rightarrow n+\pi^+$ full stars 
are data from \protect\cite{FF71}; the dashed lines are predictions of the 
present model.}
\end{figure}

\begin{figure}
\caption{ Dependence of the variable $\vec {Q^2}$ on $\theta_{\pi}$. The thin 
(thick) 
lines correspond to $W=1.2~(1.137)$ 
GeV,  $-k^2$=0.5  (GeV/c)$^2$ (full line) $-k^2$=1 (GeV/c)$^2$  (dashed line) 
$-k^2$=1.5 (GeV/c)$^2$  (dotted line) $-k^2$=2 (GeV/c)$^2$ (dashed-dotted line)}
\end{figure}


\begin{figure}
\caption{ $\theta_{\pi}$-dependence of the different contributions to the 
exclusive differential cross section for $d(e,e\pi^0)d$,  $H_{xx}+ H_{yy}$, 
$H_{xx}- H_{yy}$, $H_{zz}$ and $H_{xz}+H_{zx}$ at $W$=1.137 GeV, for different 
mechanisms contributing to the elenmentary process $\gamma^*+N\to N+\pi^0$
$\Delta-$contribution only (dotted line), $\Delta+s+u$ contributions 
(dashed-dotted line), $\Delta+s+u-\omega$ (dashed line) $\Delta+s+u+\omega$ 
(full line)}
\end{figure}
\begin{figure}
\caption{ Same as Fig. 10, but for $W=1.2$ GeV.}
\end{figure}
\begin{figure}
\caption{$\theta_{\pi}$-dependence of the ratio $r_{\pm}(\cos\theta_\pi)$  for 
$W=1.137$ GeV, with dipole $G_{Ep}$, and Paris wave function.
The $r_+$ contribution is reported for $Q^2$=0.5  (GeV/c)$^2$ (full line) and 
for 
$Q^2$=2 (GeV/c)$^2$ (dotted line). The $r_-$ contribution is reported for 
$Q^2$=0.5  
(GeV/c)$^2$ (dashed line) and for $Q^2$=2 (GeV/c)$^2$ (dashed-dotted line).} 
\end{figure}
\begin{figure}
\caption{ Same as Fig. 12, but for $W=1.2$ GeV.}
\end{figure}
\begin{figure}
\caption{ Sensitivity of the four observables to the deuteron wave function, for 
Paris (full line) and Bonn(dashed line) potentials.}
\end{figure}

\begin{figure}
\caption{ $\theta_{\pi}$ dependence of the four observables for different 
parametrization of the electromagnetic form factor of the 
$\gamma^*\pi\omega$-vertex and electric form factor of the proton at $W$=1.137 
GeV, $-k^2$=0.5 (GeV/c)$^2$ and hard form factor $G_{\omega}$:
Paris potential and $soft$ $G_{Ep}$ (full line), Paris potential and $dipole$ 
$G_{Ep}$ (line), Bonn Potential and $dipole$ $G_{Ep}$ (dotted line), Bonn 
Potential and $soft$ $G_{Ep}$ (dashed-dotted line).}
\end{figure}
\begin{figure}
\caption{ Same as Fig. 15, but for soft form factor $G_{\omega}$.}
\end{figure}
\begin{figure}
\caption{ Same as Fig. 15, but for  $-k^2$=2.0 (GeV/c)$^2$
and  hard form factor $G_{\omega}$.}
\end{figure}
 
\begin{figure}
\caption{ Same as Fig. 15, but for  $-k^2$=2.0 (GeV/c)$^2$
and   soft form factor $G_{\omega}$.}
\end{figure}
\begin{figure}
\caption{ Two-dimensional plot of the  $-k^2$ and $W$ -dependences of the 
longitudinal $H_\ell$ and transversal $H_t$ contributions to the inclusive 
differential cross section for $d(e,e')\pi^0d$ ($H_\ell$ and $H_t$ are 
dimensionless numbers).}
\end{figure}   
%
%
 

\begin{table*}
\begin{tabular}{c|c|c|c|c} 
 & $d$ &  $\vec d$ & $\stackrel{\rightarrow}{\stackrel{\rightarrow}{ d}}$ & sum 
\\ \hline 
$e$  & 4(+) & 8(-)&  16(+) & 28 	\\ \hline 
$\vec e$  & 1(-) & 5(+)&  7(-) & 13 	\\ \hline 
sum  & 5 & 13&  23 & 41	\\
\end{tabular}
\caption{Classification of Structure Functions. The sign $\pm$ denotes T-even 
and T-odd SF's}
\label{tab2}
\end{table*}
\newpage

\newpage
\begin{center}
\mbox{\epsfxsize=16.cm\leavevmode \epsffile{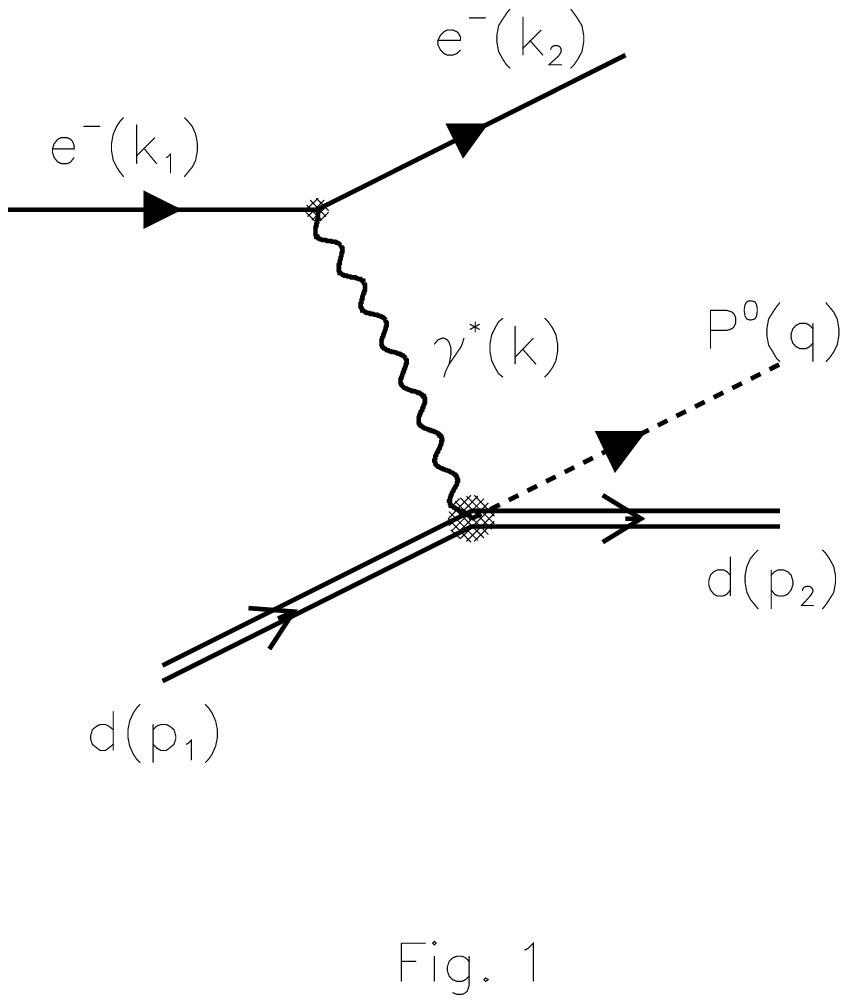}}
\end{center}

\newpage
\begin{center}
\mbox{\epsfxsize=14.cm\leavevmode \epsffile{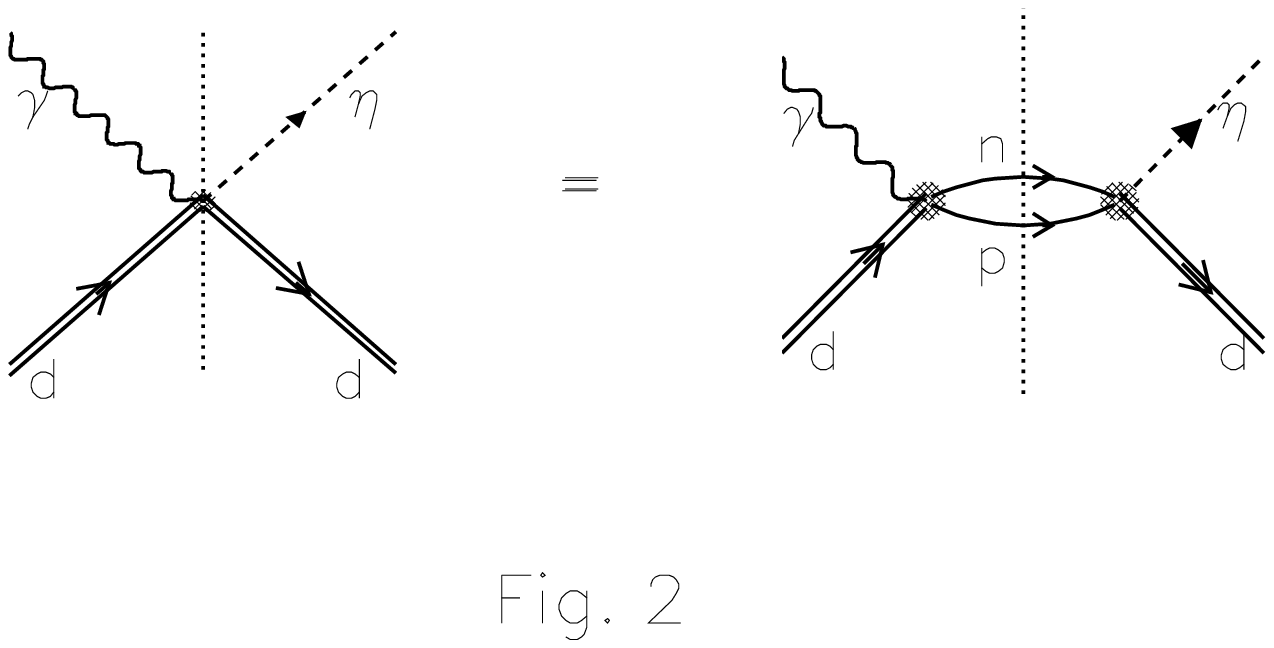}}
\end{center}

\newpage
\begin{center}
\mbox{\epsfxsize=14.cm\leavevmode \epsffile{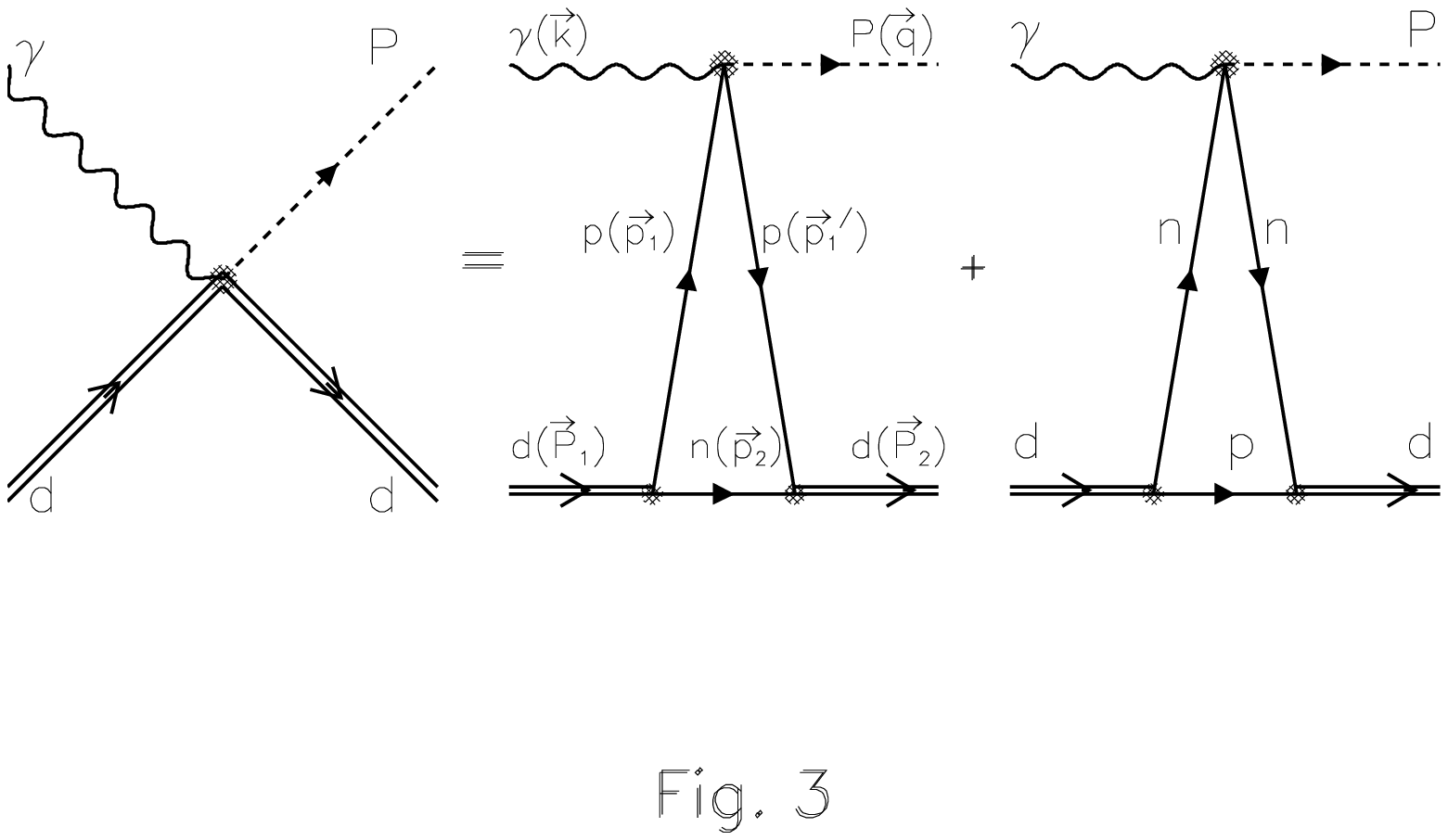}}
\end{center}

\newpage
\begin{center}
\mbox{\epsfxsize=12.cm\leavevmode \epsffile{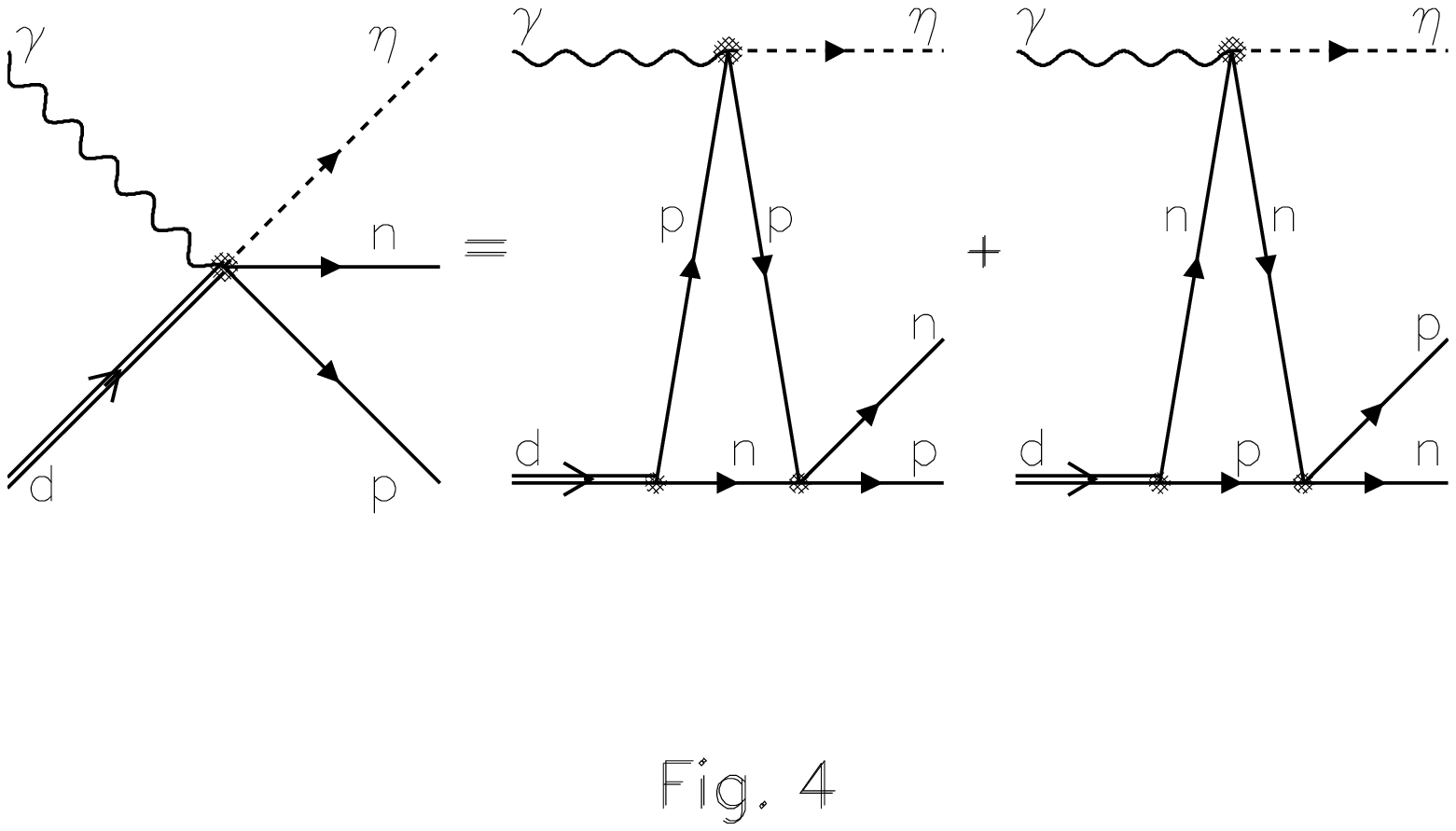}}
\end{center}

\newpage
\begin{center}
\mbox{\epsfxsize=12.cm\leavevmode \epsffile{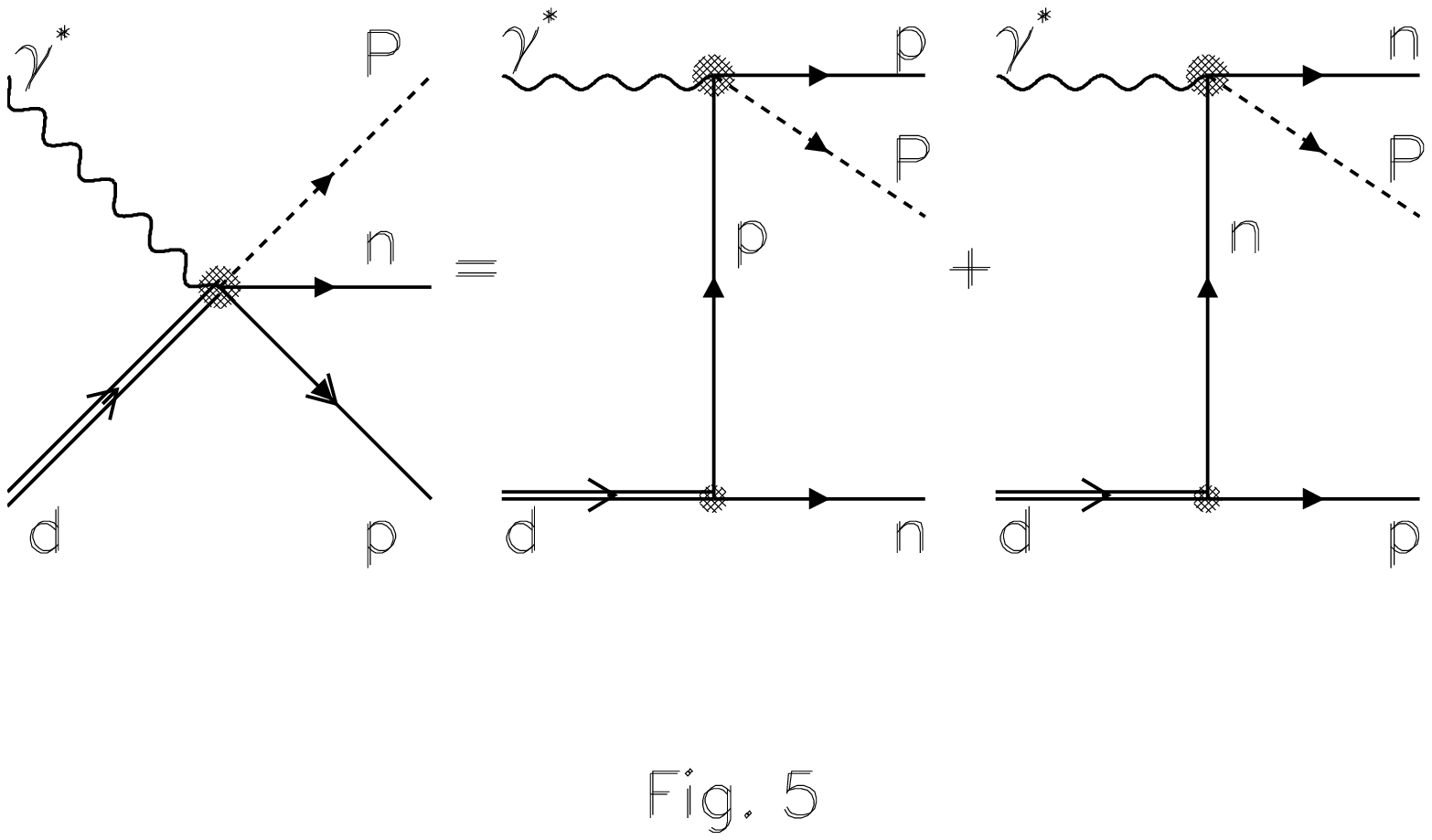}}
\end{center}

\newpage
\begin{center}
\mbox{\epsfxsize=12.cm\leavevmode \epsffile{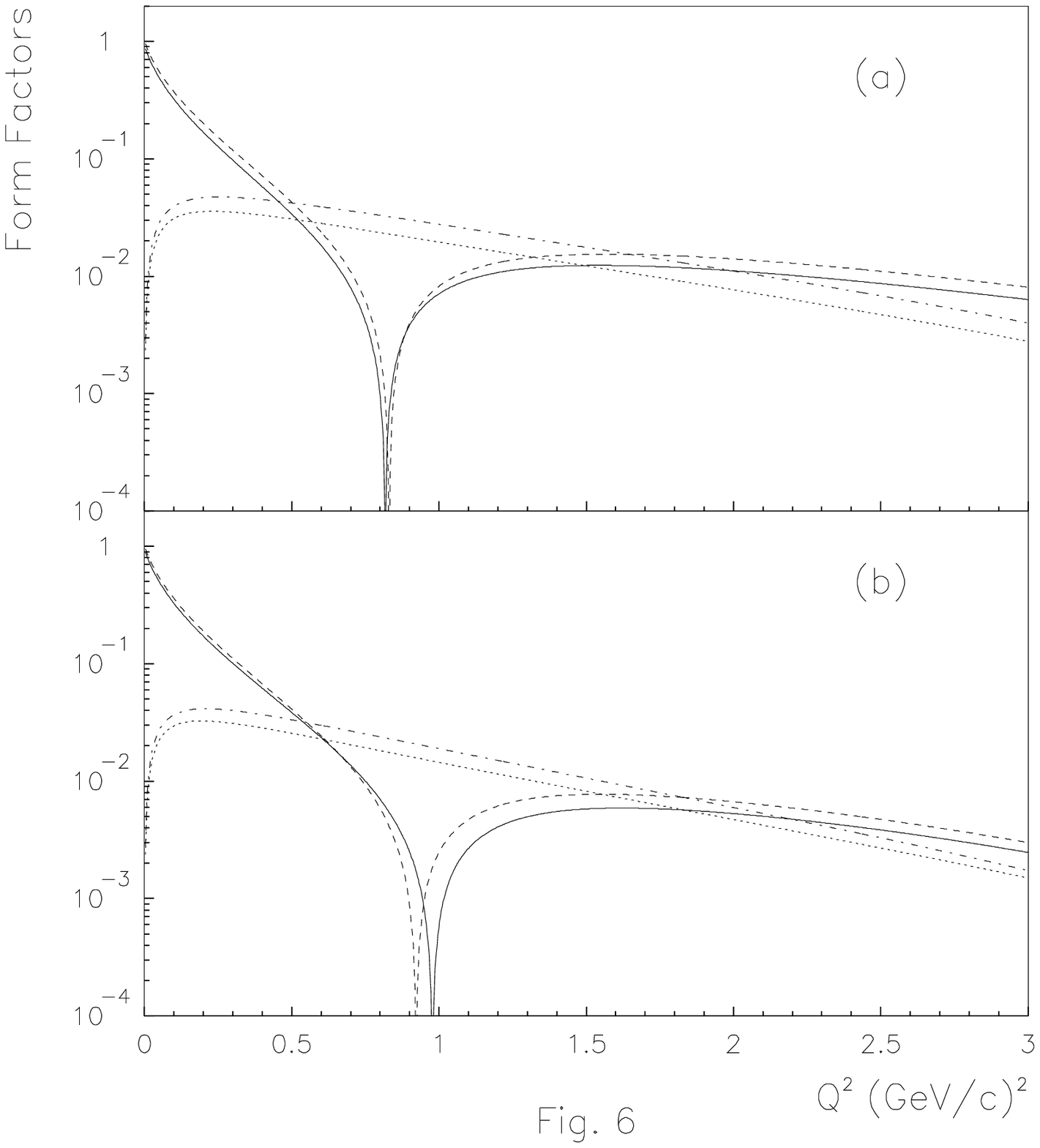}}
\end{center}

\newpage
\begin{center}
\mbox{\epsfxsize=10.cm\leavevmode \epsffile{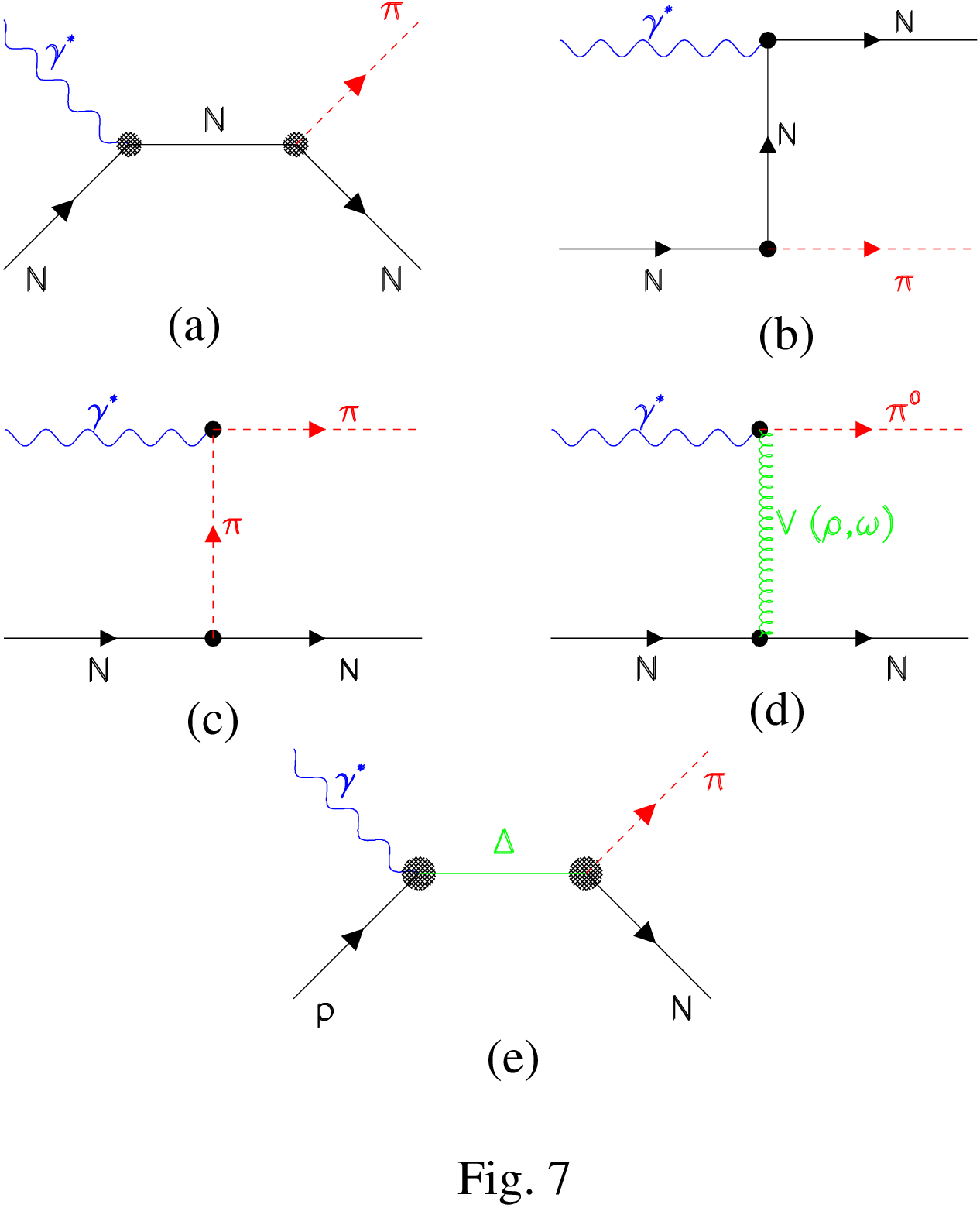}}
\end{center}

\newpage
\begin{center}
\mbox{\epsfxsize=12.cm\leavevmode \epsffile{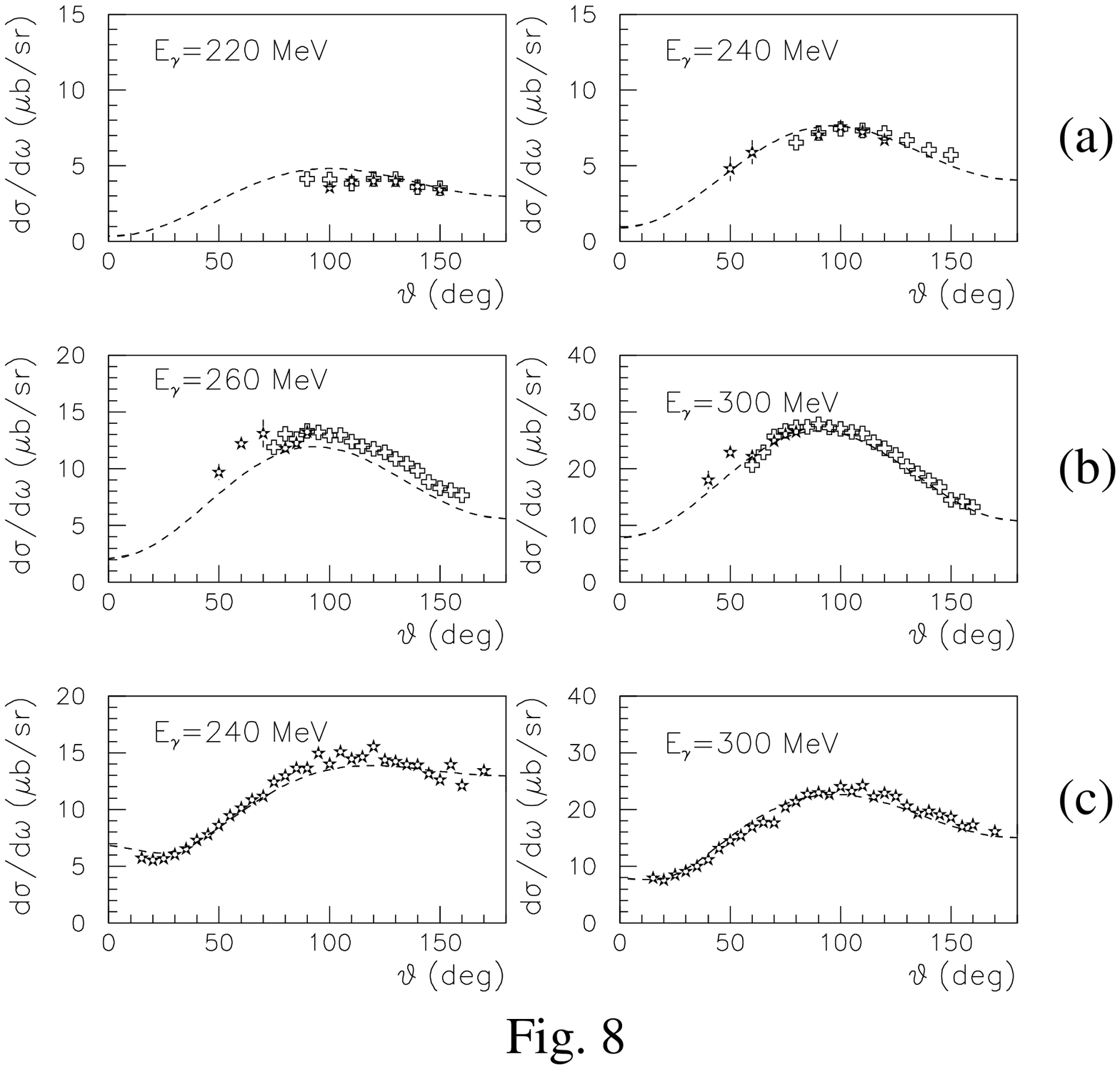}}
\end{center}

\newpage
\begin{center}
\mbox{\epsfxsize=12.cm\leavevmode \epsffile{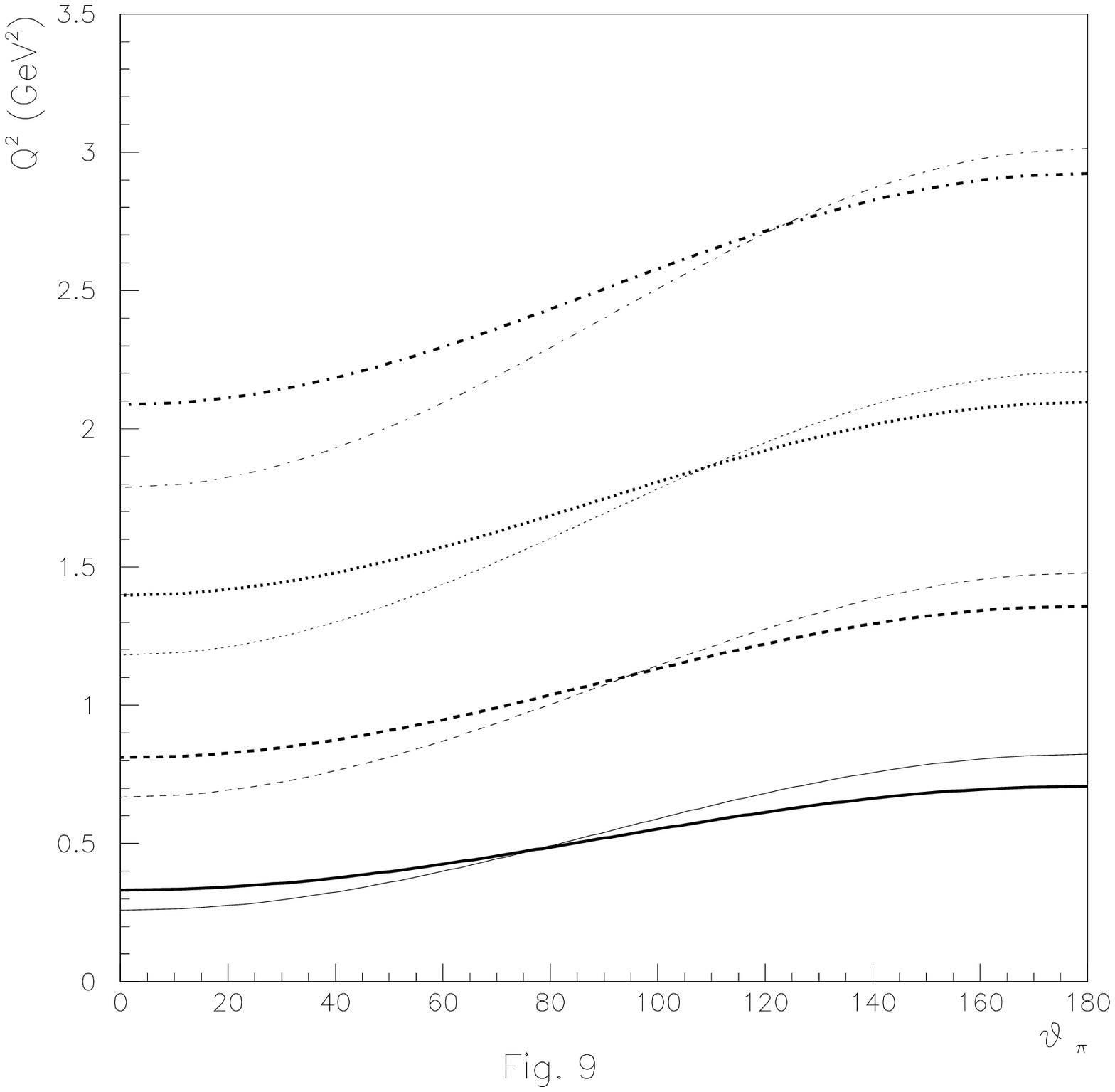}}
\end{center}

\newpage
\begin{center}
\mbox{\epsfxsize=12.cm\leavevmode \epsffile{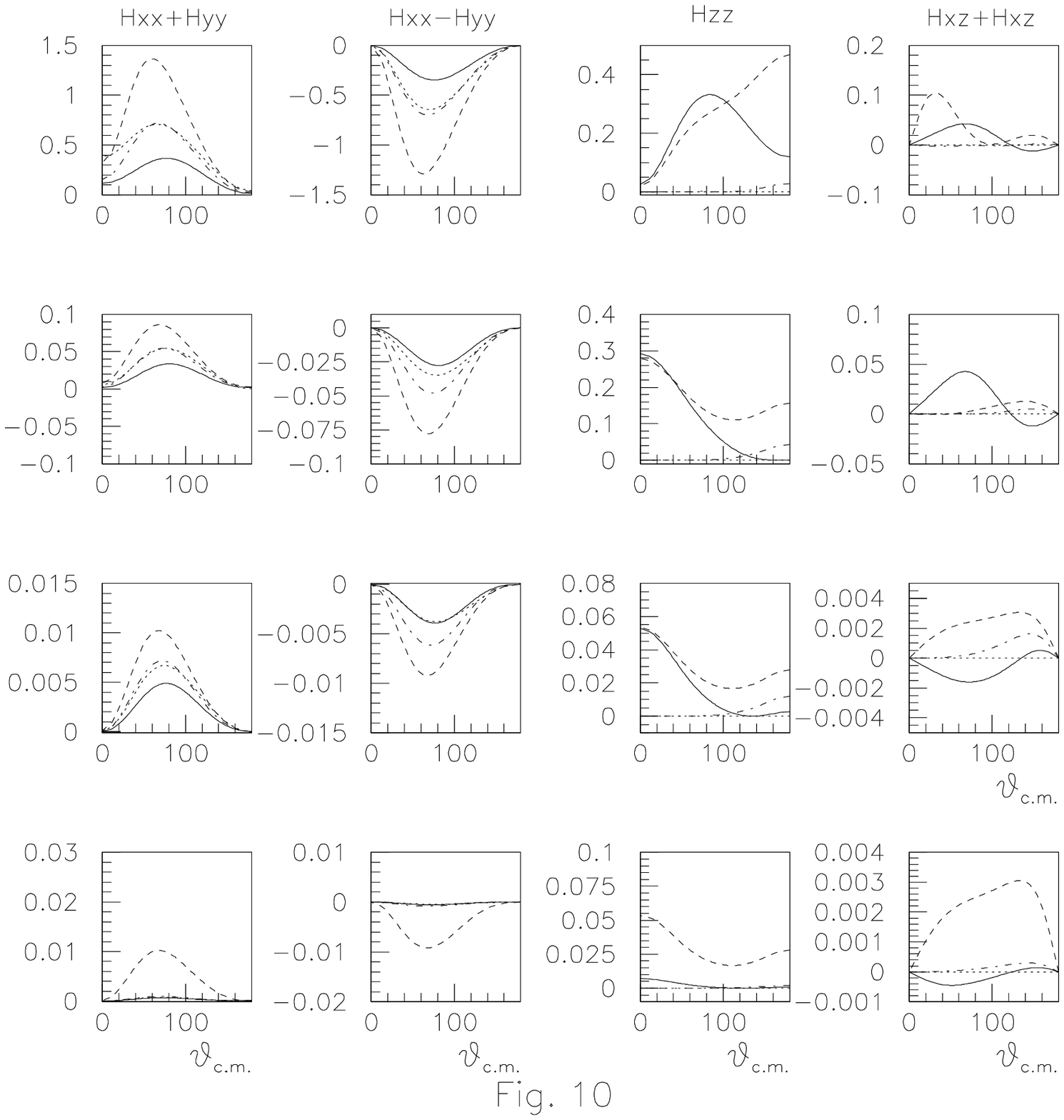}}
\end{center}

\newpage
\begin{center}
\mbox{\epsfxsize=12.cm\leavevmode \epsffile{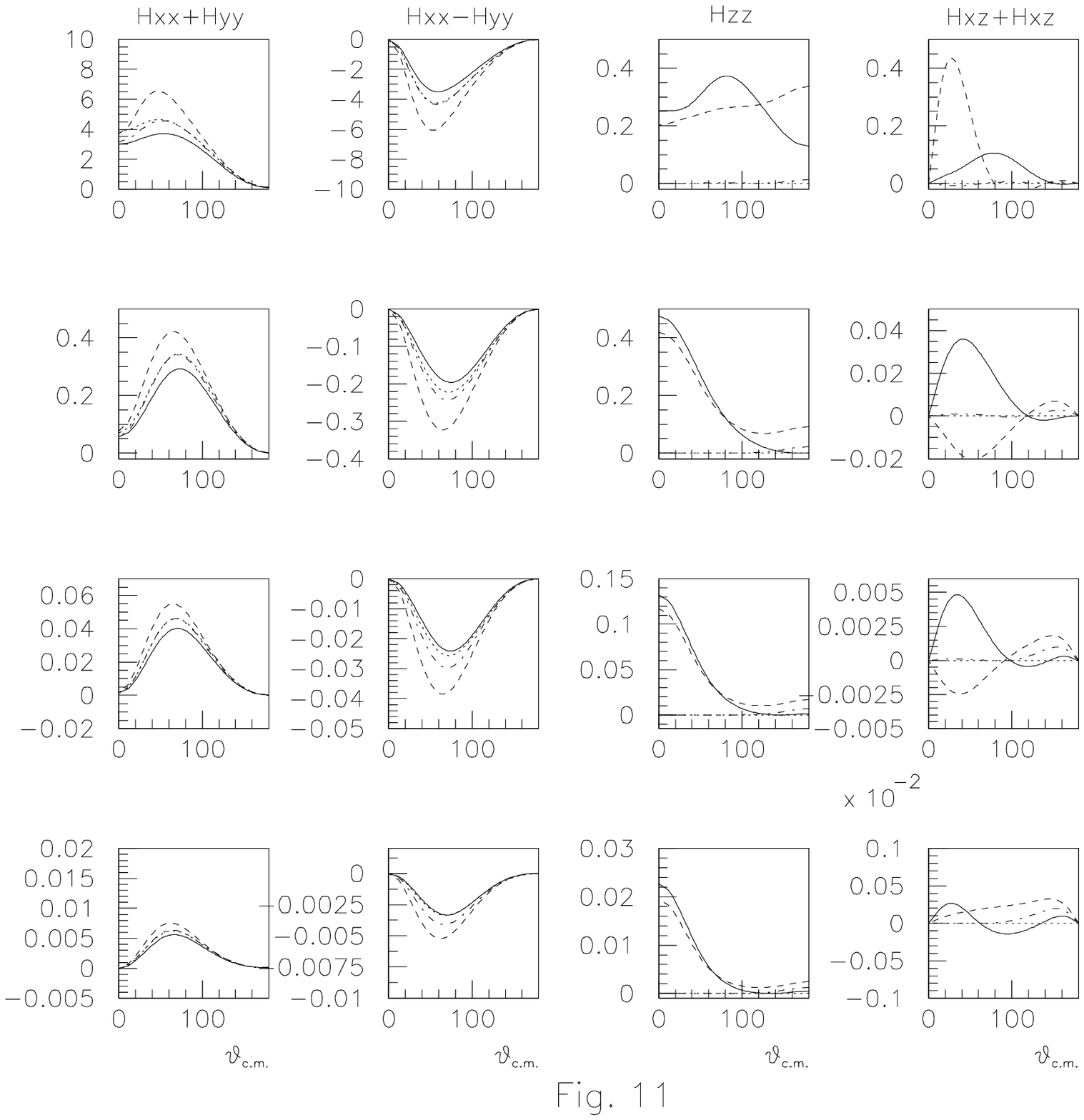}}
\end{center}

\newpage
\begin{center}
\mbox{\epsfxsize=12.cm\leavevmode \epsffile{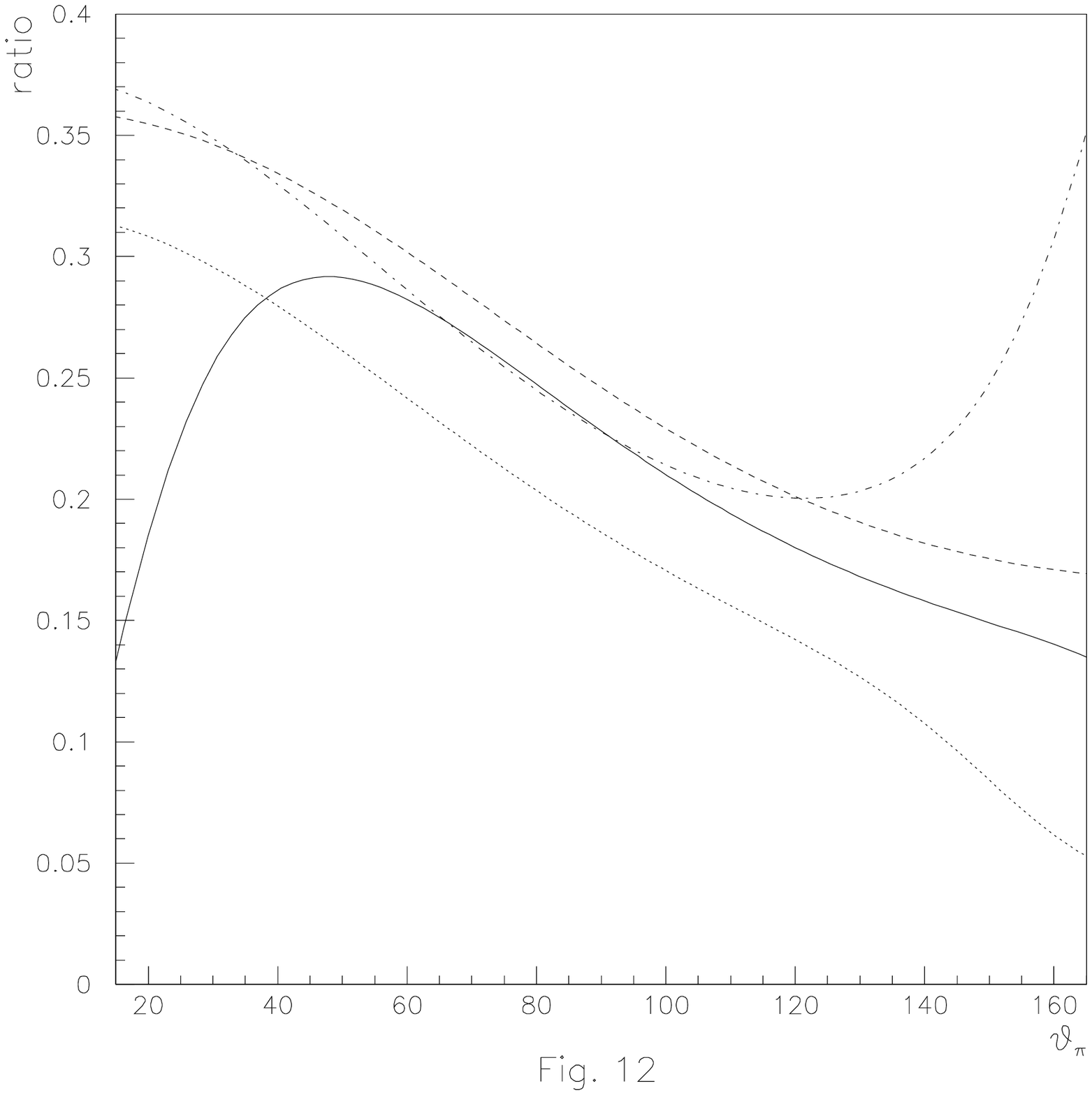}}
\end{center}

\newpage
\begin{center}
\mbox{\epsfxsize=12.cm\leavevmode \epsffile{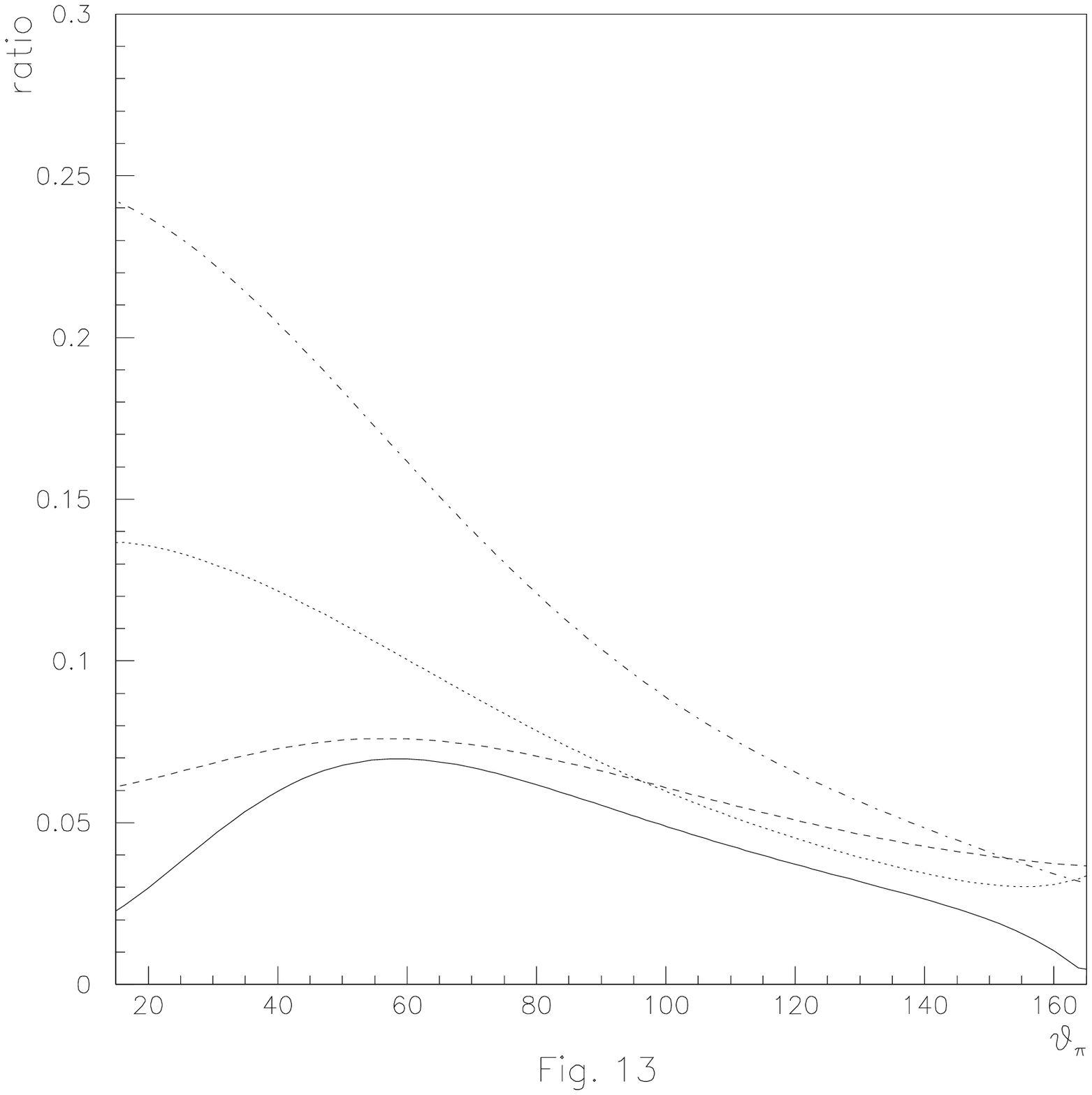}}
\end{center}

\newpage
\begin{center}
\mbox{\epsfxsize=12.cm\leavevmode \epsffile{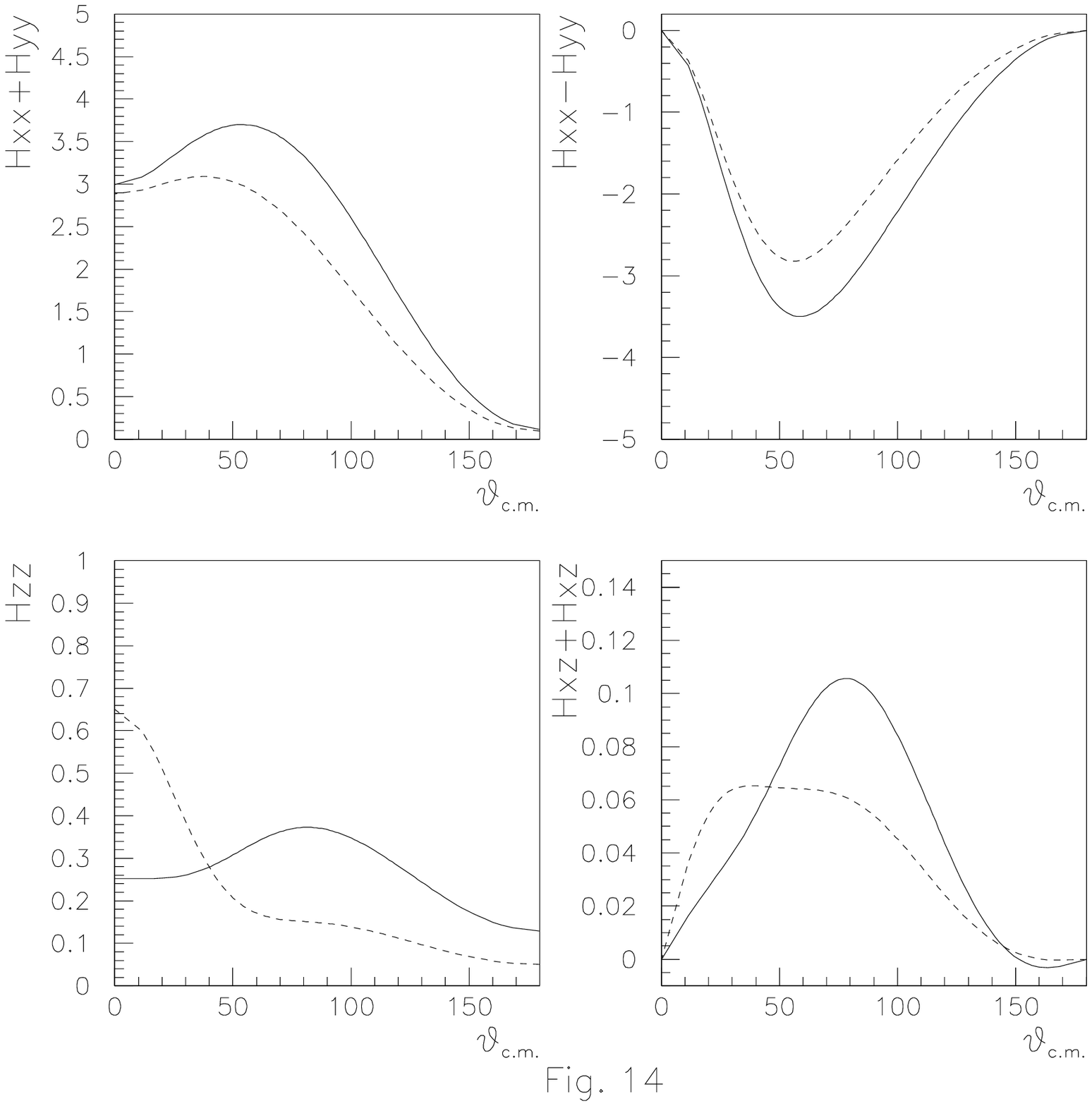}}
\end{center}

\newpage
\begin{center}
\mbox{\epsfxsize=12.cm\leavevmode \epsffile{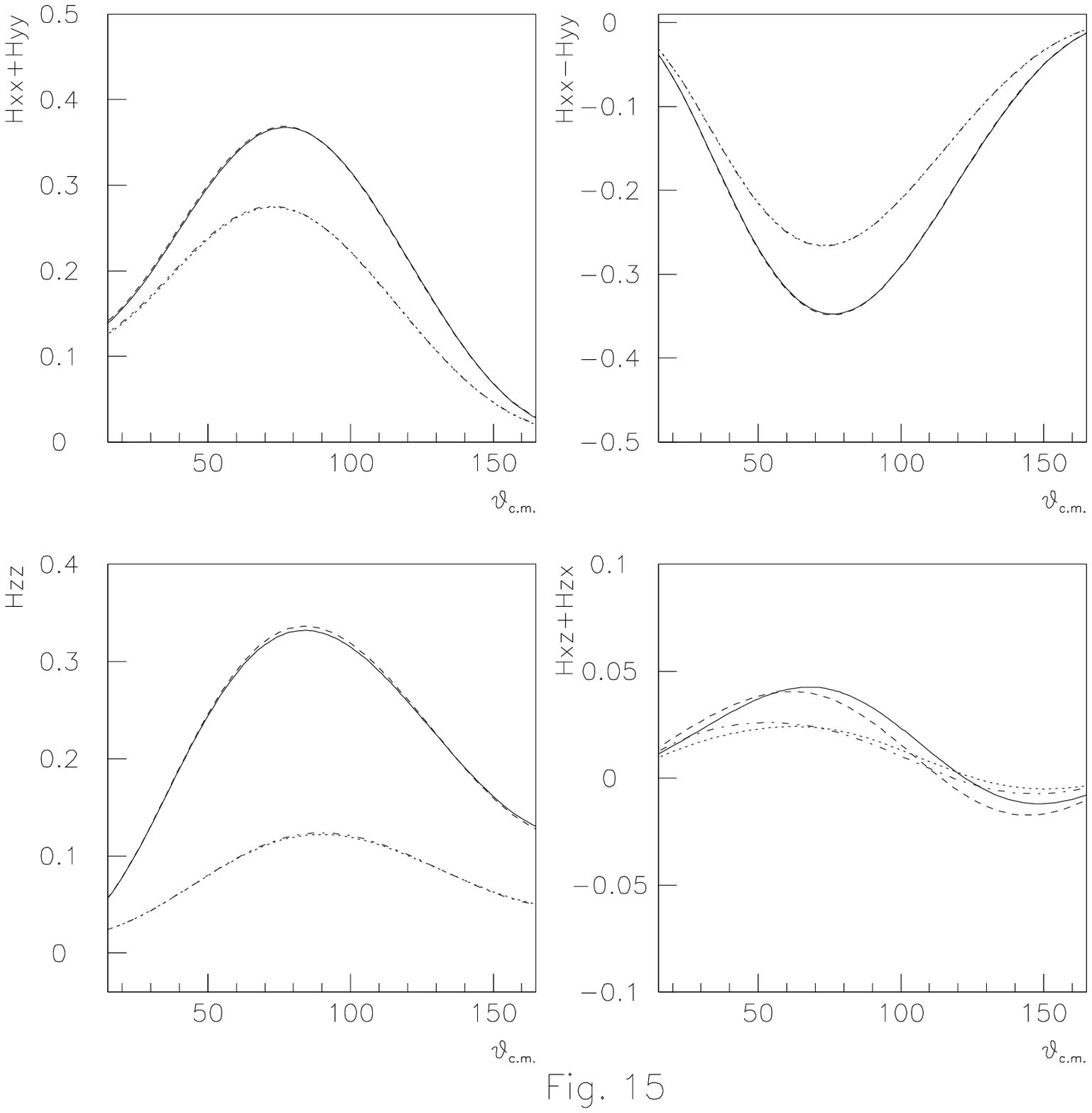}}
\end{center}

\newpage
\begin{center}
\mbox{\epsfxsize=12.cm\leavevmode \epsffile{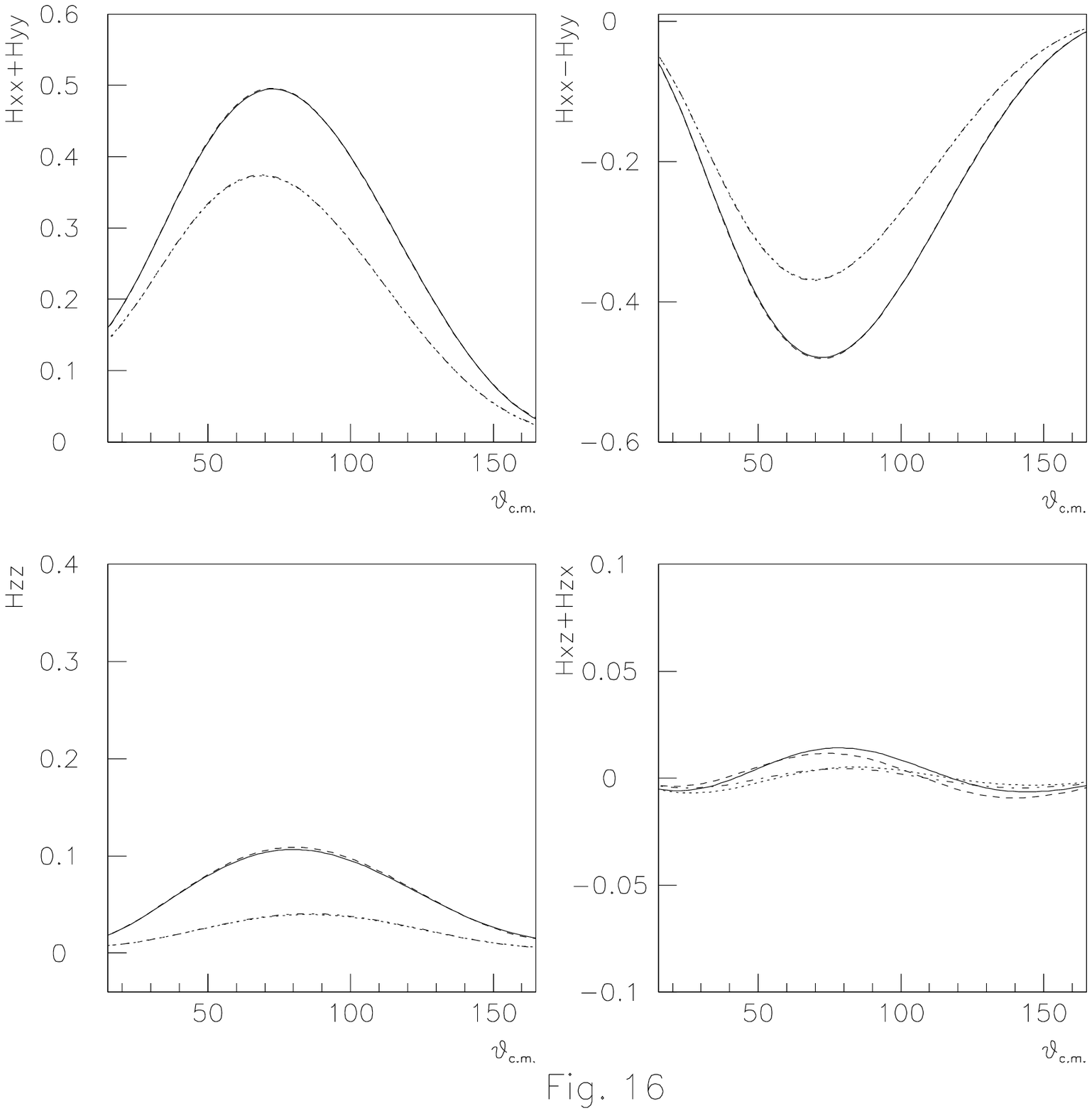}}
\end{center}

\newpage
\begin{center}
\mbox{\epsfxsize=12.cm\leavevmode \epsffile{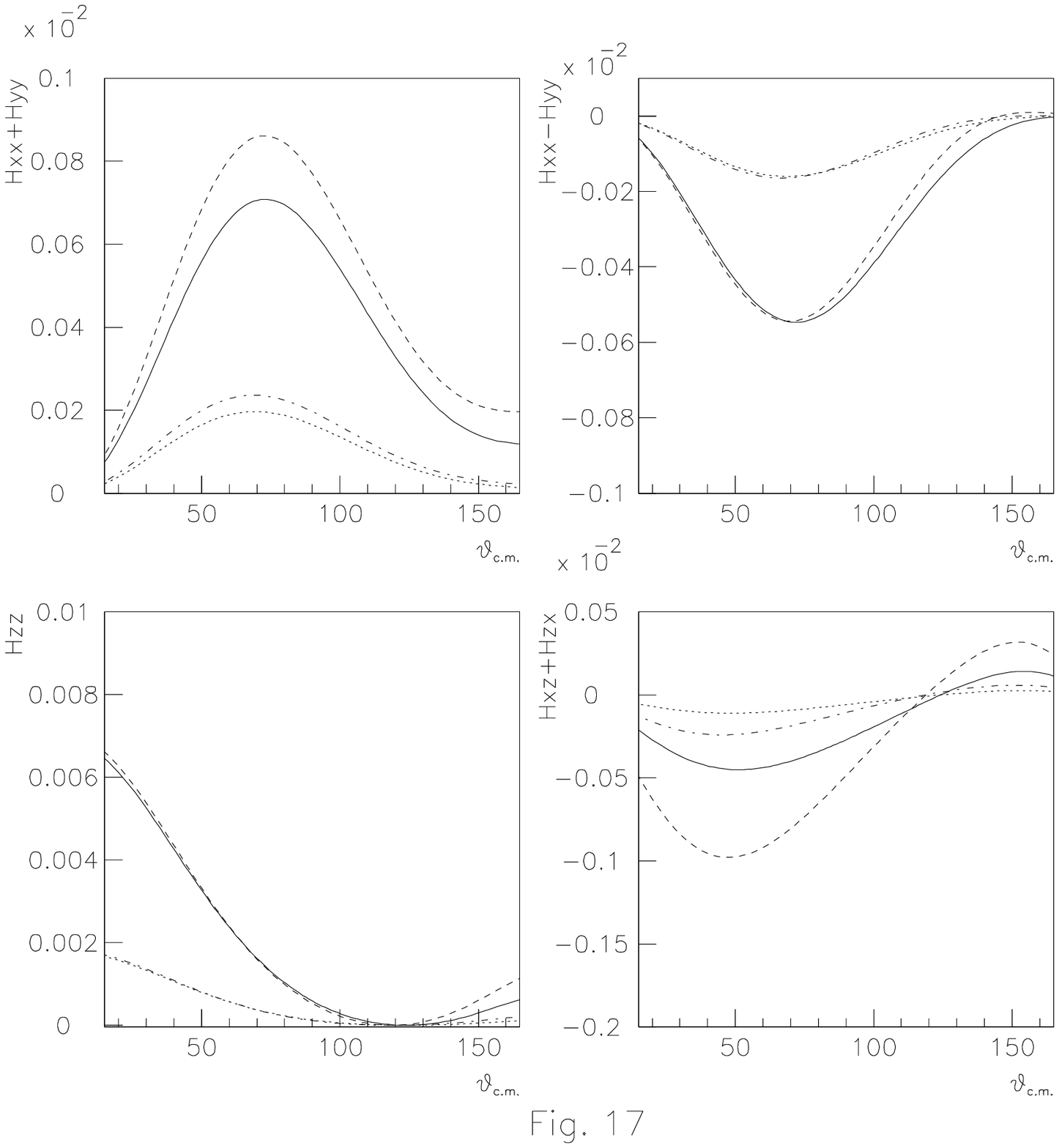}}
\end{center}
\newpage
\begin{center}
\mbox{\epsfxsize=12.cm\leavevmode \epsffile{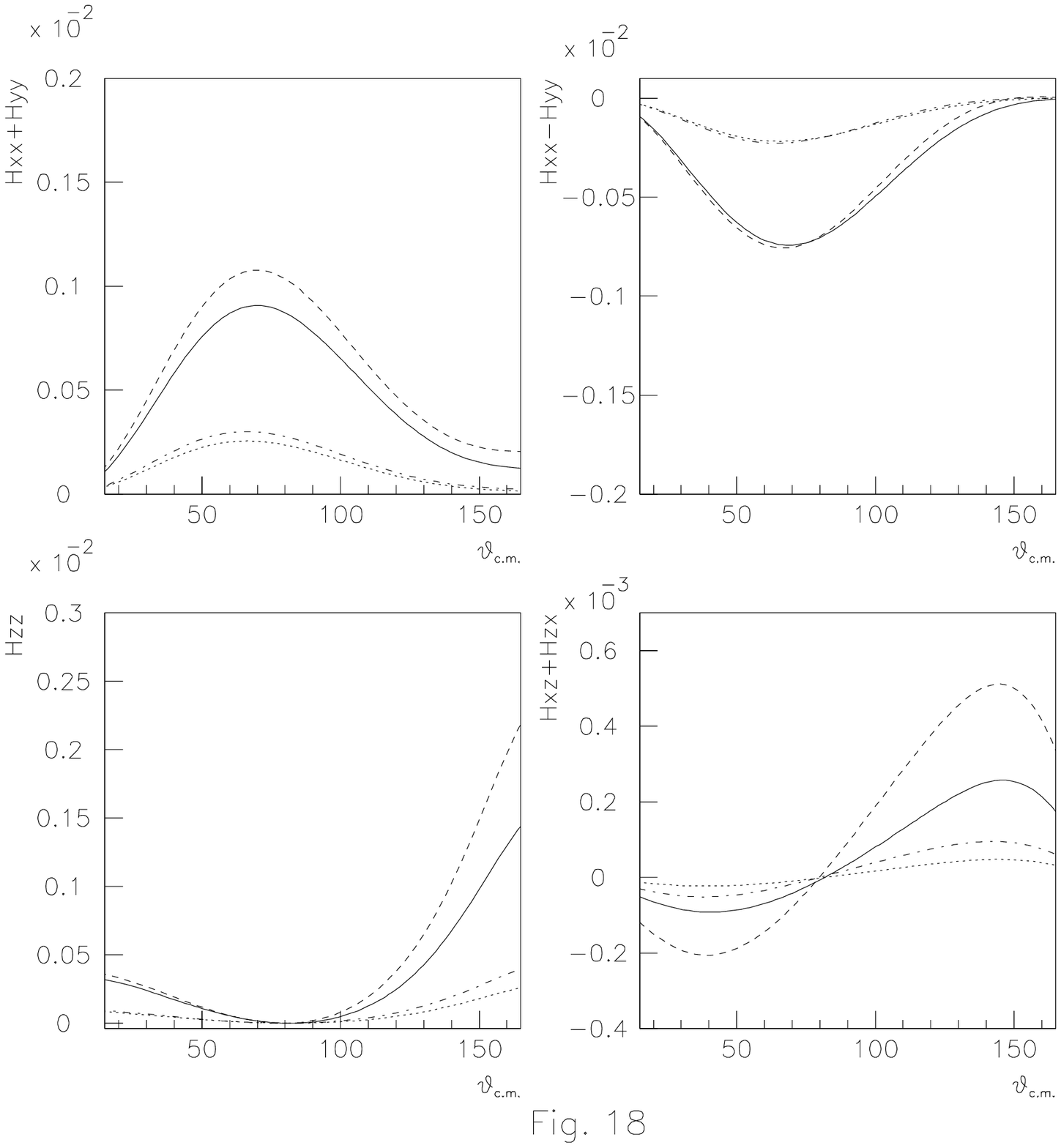}}
\end{center}
\newpage
\begin{center}
\mbox{\epsfxsize=12.cm\leavevmode \epsffile{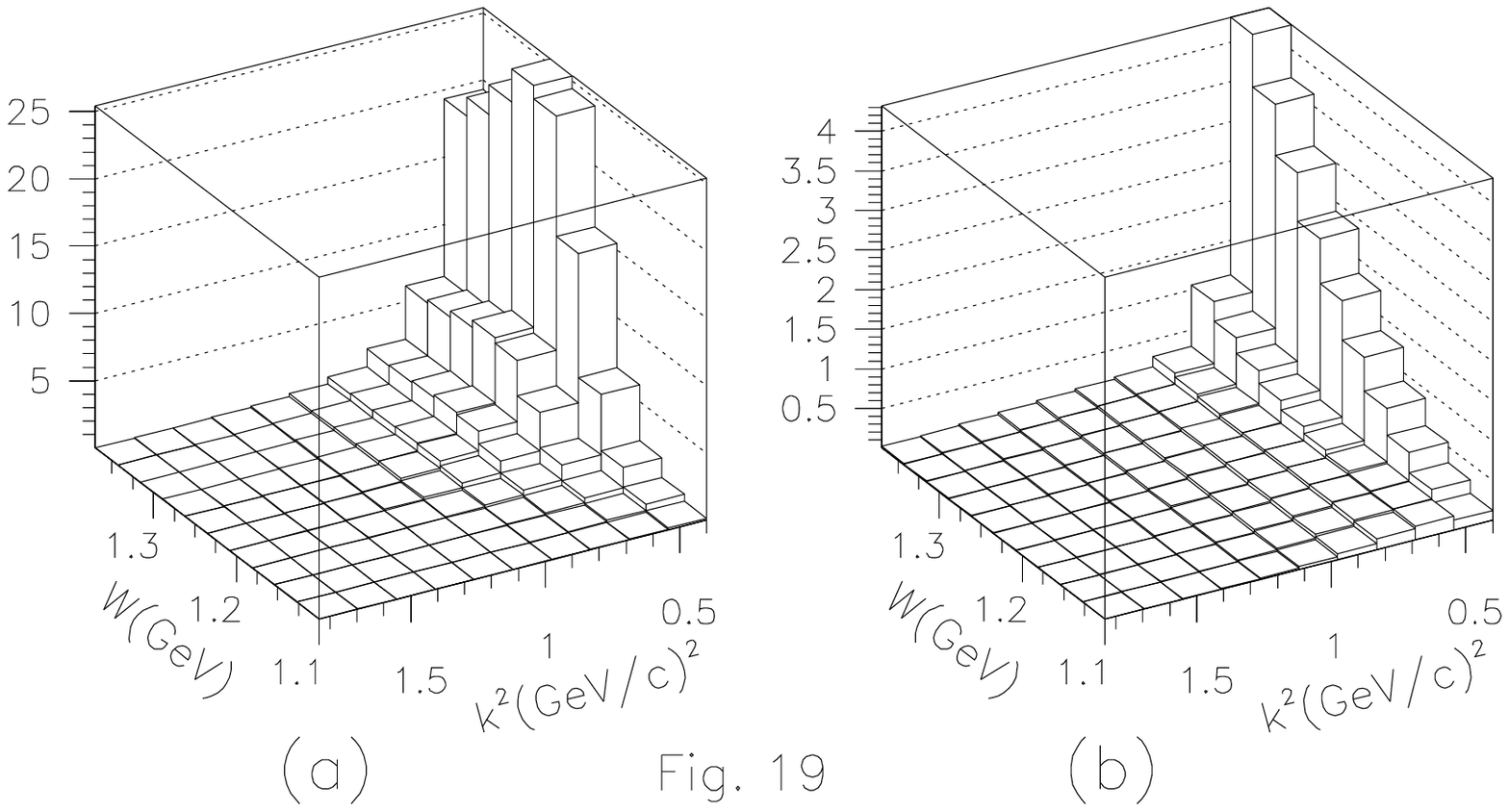}}
\end{center}
\end{document}